\documentclass[aps, twocolumn, superscriptaddress, longbibliography]{revtex4-2}
\usepackage[english]{babel}
\usepackage{graphicx}
\usepackage{stmaryrd}
\usepackage{amsmath, amsfonts, amssymb}
\usepackage{mathrsfs}
\usepackage{braket}
\usepackage{color}
\usepackage{xspace}
\usepackage{amscd}
\usepackage{latexsym}
\usepackage{mathtools}
\usepackage{diagbox}
\usepackage{bm}
\usepackage{tikz}
\usepackage[caption=false]{subfig}
\usetikzlibrary{matrix}
\usetikzlibrary{shapes.geometric}
\usetikzlibrary{calc}
\usepackage[
colorlinks=true,
urlcolor=blue,
citecolor=blue,
linkcolor=blue,
hyperfootnotes=false]{hyperref}
\usepackage{cleveref}

\usepackage[ruled,vlined]{algorithm2e}
\SetKwProg{Fn}{Function}{}{}

\usepackage[mode=buildnew]{standalone}%

\definecolor{stabblue}{HTML}{D0D0FE}
\definecolor{stabred}{HTML}{FED0D0}

\newcommand\gap{0.5}
	\newcommand\bl{0}
	\newcommand\br{0.5}

	\newcommand\w{0.15}
	\newcommand\gsize{0.15}
	\newcommand\zlcol{black!40!blue}
	
	\newcommand\zcol{blue}
	
	\newcommand\op{0.2}
	\newcommand{\whisker}[4]{
	\foreach \k in {-1,...,1}{
		\draw[#3] ($(#1) + (#2) + (0,\k*0.03) $) -- (#1);
	}
	}
	\newcommand{\pairstab}[7]{
		\node [circle,draw,fill=#3,#3,scale=\gsize] at (#4,\gap*#1+#7) {};
		\node [circle,draw,fill=#3,#3,scale=\gsize] at (#5,\gap*#1+#7) {};
		\node[draw=#3,minimum width={(#5-#4+2*\w)*1cm},minimum height=2*\w cm, fill=#3,opacity=\op,rounded corners] (#6#1) at ({(#4+#5)/2},\gap*#1+#7) {};
		\whisker{#6#1}{{-(#5-#4+2*\w)},0}{#2}{1};
		\whisker{#6#1}{{(#5-#4+2*\w)},0}{#2}{1};
	}
	\newcommand{\singlestab}[7]{
		\node [circle,draw,fill=#3,#3,scale=\gsize] at (#4,\gap*#1) {};
		\node [circle,draw,fill=#3,#3,scale=\gsize] at (#5,\gap*#1) {};
		\node[draw=#3,minimum width={(2*\w)*1cm},minimum height=2*\w cm, fill=#3,opacity=\op,rounded corners] (#6#1) at (#4,\gap*#1) {};
		\node[draw=#3,minimum width={(2*\w)*1cm},minimum height=2*\w cm, fill=#3,opacity=\op,rounded corners] (#7#1) at (#5,\gap*#1) {};
		\whisker{#6#1}{{-(#5-#4)},0}{#2}{1};
		\whisker{#7#1}{{(#5-#4)},0}{#2}{1};
	}

\begin{document}
 \title{Subsystem codes with high thresholds by gauge fixing and reduced qubit overhead}
	
	\author{Oscar Higgott}
	\email{oscar.higgott.18@ucl.ac.uk}
	\author{Nikolas P. Breuckmann}
	\email{n.breuckmann@ucl.ac.uk}
	\affiliation{Department of Physics \& Astronomy, University College London, WC1E 6BT London, United Kingdom}
	
	\date{\today}
	
	\begin{abstract}
	We introduce a technique that uses gauge fixing to significantly improve the quantum error correcting performance of subsystem codes.
	By changing the order in which check operators are measured, valuable additional information can be gained, and we introduce a new method for decoding which uses this information to improve performance.
	Applied to the subsystem toric code with three-qubit check operators, we increase the threshold under circuit-level depolarising noise from~$0.67\%$ to~$0.81\%$.
	The threshold increases further under a circuit-level noise model with small finite bias, up to $2.22\%$ for infinite bias.
	Furthermore, we construct families of finite-rate subsystem LDPC codes with three-qubit check operators and optimal-depth parity-check measurement schedules.
	To the best of our knowledge, these finite-rate subsystem codes outperform all known codes at circuit-level depolarising error rates as high as~$0.2\%$, where they have a qubit overhead that is~$4.3\times$ lower than the most efficient version of the surface code and~$5.1\times$ lower than the subsystem toric code.
	Their threshold and pseudo-threshold exceeds $0.42\%$ for circuit-level depolarising noise, increasing to $2.4\%$ under infinite bias using gauge fixing.
	\end{abstract}

	\maketitle

	\section{Introduction}\label{sec:introduction}
	
	The realization of scalable quantum computing depends on our ability to correct errors which arise due to inevitable interactions between the device and the environment.
	These errors can be corrected by introducing redundancy in the form of quantum error correcting codes.
	The most widely studied quantum error correcting code, both theoretically and experimentally, is the surface code~\cite{dennis2002topological}, which has a high tolerance for realistic circuit-level noise and uses four-qubit measurements that are geometrically local in two dimensions.
	
	Despite these advantages, the surface code has several shortcomings. 
	Firstly, it is estimated that thousands of physical qubits will be required to encode each logical qubit for fault-tolerant quantum computing in a noise regime of practical interest~\cite{fowler2012surface}.
	While there has been significant progress in the construction of families of codes, called quantum \textit{low-density parity-check} (LDPC) codes, that have improved theoretical parameters relative to the surface code~\cite{breuckmann2021ldpc,tillich2013quantum,freedman2002z2,breuckmann2016constructions,guth2014quantum,breuckmann2020single}, none of these codes have been shown to have a lower qubit overhead than the surface code once circuit-level noise is taken into account.
	Secondly, even weight-four check operator measurements can be too large for some architectures. In some superconducting qubit architectures, for example, the degree of the surface code interaction graph can lead to frequency collisions~\cite{chamberland2020topological}.
	Finally, physical error rates observed experimentally in devices are still above the threshold~\cite{chen2021exponential}, and the standard implementation of the surface code is not well suited to handle biased noise models that can arise in some physical systems~\cite{puri2020bias}.
	As a result, improving the tolerance of the surface code to biased noise models is an active area of research~\cite{tuckett2020fault,bonilla2020xzzx,xu2019high,stephens2013high}.
	
	In this work, we tackle all three of these problems by introducing new decoding techniques and constructions for subsystem codes. 
	Most quantum codes considered in the literature are stabiliser codes, which are defined in terms of a set of Pauli operators~\cite{gottesman1997stabilizer}.
	Subsystem codes are a slight generalisation of stabiliser codes where only a subset of the available encoded degrees of freedom are used~\cite{poulin2005stabilizer}.
	They can simplify the measurements which are part of the error correction procedure by reducing the number of physical qubits involved~\cite{aliferis2007subsystem}, or by enabling bare-ancilla fault-tolerance even when the check weights are large~\cite{li2018direct}.
	Furthermore, subsystem codes allow for a procedure called \emph{gauge fixing} which is useful to manipulate the encoded quantum information.
	Gauge fixing effectively allows us to change the code mid-computation, and in~\cite{paetznick2013universal} the authors exploit this to switch between codes which have complementary sets of logical operations.
	
	These advantages have motivated experimentalists to pursue subsystem codes for implementing fault-tolerant quantum computation. This includes IBM, who plan to implement the heavy-hexagon subsystem code~\cite{chamberland2020topological} to reduce frequency collisions in their superconducting quantum processors~\cite{jurcevic2020demonstration}. Notably, the Bacon-Shor subsystem code~\cite{bacon2006operator} has recently been implemented experimentally in a trapped-ion architecture~\cite{egan2020fault}, where the fidelity of the encoded logical operations exceeded that of the entangling physical operations used to implement them.

	However, subsystem codes have usually had lower thresholds, an issue which can be attributed to their higher weight stabilisers.
	They have also typically had a larger qubit overhead, since some logical qubits are not used to encode information.
	We introduce constructions and decoding techniques that instead demonstrate that subsystem codes can be used to \textit{increase} the threshold and \textit{reduce} the qubit overhead. The decoding technique we introduce, called schedule-induced gauge fixing, improves the error correcting performance of a wide class of subsystem codes, especially under biased noise models.
	By changing the order in which check operators are measured, valuable additional information can be gained, and we introduce a new method for decoding which uses this information to improve performance.
	In previous work, the Bacon-Shor code has been used as a template to construct elongated compass codes, which can be tailored to biased noise models~\cite{li20192d}.
	However, this requires changing interactions at the hardware level, as well as measuring high weight stabilisers directly, since elongated compass codes are not subsystem codes themselves.
	In contrast, our technique can be implemented entirely in software, and only requires measuring the low-weight gauge operators of the code.	
	In essence, schedule-induced gauge fixing allows us to switch repeatedly between different codes such that more information can be inferred about potential errors.
	
	We also reduce the qubit overhead for quantum error correction by introducing a construction for subsystem codes that encode a number of logical qubits $k$ proportional to the number of physical qubits $n$, while using only three-qubit check operators.
	These codes are derived from hyperbolic tessellations, and we use the symmetry group of the tessellation to derive quantum circuits for measuring the check operators that use only four time steps, which is optimal.
	From simulating their performance in circuit-level depolarising noise, we find that these finite-rate subsystem codes have a qubit overhead that is~$4.3\times$ lower than the most efficient version of the surface code for error rates as high as 0.2\%, which is a noise regime often considered for practical surface code quantum computing~\cite{gidney2019factor}.
	 The potential advantages of quantum LDPC codes have previously only been shown under a simplistic phenomenological noise model~\cite{breuckmann2017hyperbolic}, in some cases at very low error rates~\cite{li2020numerical}.
	Once circuit-level noise is taken into account, the potential reduction in qubit overhead can be lost~\cite{conrad2018small}.
	Therefore, to the best of our knowledge, the results for our finite-rate subsystem codes are the first demonstration of a quantum code outperforming the surface code in a practical regime of circuit-level depolarising noise.	
	
	In \Cref{sec:preliminaries} we review the stabiliser formalism, subsystem codes and gauge fixing, before reviewing the subsystem surface code in \Cref{sec:2d_construction} in the context of our more general subsystem code construction. We introduce our construction for finite-rate subsystem LDPC codes in \Cref{sec:subsystem_hyperbolic_codes}, where we analyse their properties and show how to construct efficient stabiliser measurement circuits for them. In \Cref{sec:gauge_fixing} we introduce schedule-induced gauge fixing, our technique for improving the quantum error correcting performance of subsystem codes. We present our numerical results in \Cref{sec:numerical_analysis}, which includes the application of schedule-induced gauge fixing to the subsystem toric code, as well as a performance analysis of our finite-rate subsystem LDPC codes. We discuss broader applications of schedule-induced gauge fixing and our constructions in \Cref{sec:broader_applications}, before concluding in \Cref{sec:conclusion}.

	\section{Preliminaries}\label{sec:preliminaries}
	
	A quantum stabilizer code is defined by an abelian subgroup~$\mathcal{S}$ of the Pauli group operating on $n$ physical qubits.
	The code space is the common $+1$-eigenspace of all elements of the stabilizer group.
	If there exists a generating set of~$\mathcal{S}$ such that each generator acts non-trivially on the physical qubits as either Pauli-$X$ or Pauli-$Z$ only then the code is called a \emph{CSS code}. 
	
	A subsystem code is a stabiliser code in which a subset of logical operators are chosen not to store information~\cite{poulin2005stabilizer}. In a subsystem code, the overall Hilbert space~$\mathcal{H}$ can be decomposed as
	\begin{equation}
	\mathcal{H}=(\mathcal{H}_\mathcal{L}\otimes\mathcal{H}_\mathcal{G})\oplus\mathcal{C}^{\perp}
	\end{equation}
	where only $\mathcal{H}_\mathcal{L}$ stores information and any operations applied only on $\mathcal{H}_\mathcal{G}$ are ignored. The Pauli operators that act trivially on $\mathcal{H}_\mathcal{L}$ form the \textit{gauge group} $\mathcal{G}$ of the code.
	The stabiliser group $\mathcal{S}$ is the center of $\mathcal{G}$ up to phase factors, $\langle iI,\mathcal{S}\rangle = Z(\mathcal{G})\coloneqq C(\mathcal{G})\cap \mathcal{G}$.
	Hence, up to phase factors, operators from $\mathcal{G}$ are either stabilisers (acting trivially on $\mathcal{H}_\mathcal{L}\otimes\mathcal{H}_\mathcal{G}$), or act non-trivially on $\mathcal{H}_\mathcal{G}$ only.
	Logical operators that act non-trivially only on $\mathcal{H}_\mathcal{L}$ are called \textit{bare} logical operators $\mathcal{L}_{bare}$, and are given by $C(\mathcal{G})\setminus\mathcal{G}$. The \textit{dressed} logical operators $\mathcal{L}_{dressed}=C(\mathcal{S})\setminus\mathcal{G}$ act non-trivially on both~$\mathcal{H}_\mathcal{L}$ and~$\mathcal{H}_\mathcal{G}$.
	A dressed logical operator is a bare logical operator multiplied by a gauge operator in $\mathcal{G}\setminus\mathcal{S}$. The distance $d$ of a subsystem code is the weight of the minimum-weight dressed logical operator, $d=\min_{P\in C(\mathcal{S})\setminus\mathcal{G}}|P|$.
	The number of physical qubits $n$, logical qubits $k$, independent stabilizer checks~$r$ and gauge qubits~$g$ are related as
	\begin{align}\label{eqn:parameter_dependency}
	n - k = r + g .
	\end{align}
	
	One advantage of introducing gauge qubits is that they can enable simpler stabiliser measurements if the generators of the gauge group $\mathcal{G}$ (the \textit{gauge generators}) have a lower weight than the generators of the stabiliser group~$\mathcal{S}$. Since $\mathcal{S}\subseteq\mathcal{G}$ the outcomes of the gauge generator measurements can be used to infer the eigenvalues of the stabilisers, provided the gauge generators are measured in the appropriate order (since $\mathcal{G}$ is generally not abelian)~\cite{suchara2011constructions, terhal2015quantum}.
	We will refer to standard stabiliser codes, where all logical qubits are used to store quantum information, as \textit{subspace} codes, to distinguish them from subsystem codes.
	
	The technique called \textit{gauge fixing}, applied to a subsystem code, consists of adding an element $g\in\mathcal{G}$ into the stabiliser group $\mathcal{S}$, as well as removing every element $h\in\mathcal{G}$ that anticommutes with~$g$. Gauge fixing was introduced by \citeauthor{paetznick2013universal} and can be useful for performing logical operations~\cite{paetznick2013universal, bombin2015gauge, yoder2017universal}, including for code deformation and lattice surgery~\cite{vuillot2019code}.
	Gauge fixing can also be used for constructing codes: both the surface code and the heavy-hexagon code~\cite{chamberland2020topological} are gauge fixings of the Bacon-Shor subsystem code~\cite{bacon2006operator}, all belonging to the larger family of 2D compass codes~\cite{li20192d}.
	These constructions are static, as the fixed gauge stays the same over time.
	In this work, we will show how gauge fixing can be used to improve the quantum error correcting performance of subsystem codes.
	Here, we consider dynamical approach to gauge fixing, i.e. in contrast to the code constructions mentioned earlier, we change which gauge degrees of freedom are being fixed over time.
	This will allow us to improve the error correction capabilities of subsystem codes.

	\section{The Subsystem Surface Code}\label{sec:2d_construction}
	
	We will now describe a method for constructing subsystem codes from hexagonal lattices, which we will see is equivalent to the subsystem toric code of \cite{bravyi2013subsystem}.
	In \Cref{sec:subsystem_hyperbolic_codes} we will generalise this construction to other tessellations to obtain subsystem \textit{hyperbolic} codes.
	
\begin{figure}[hb]
	\includegraphics{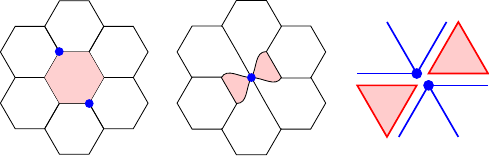}
		\caption{Left \& Middle: Merging inside a hexagonal lattice. After merging, the resulting vertex has degree six. Note that the surrounding faces are unaffected (besides being deformed). Right:~After merging we can break the $X$-check~(blue) and $Z$-check~(red) into two pairs of operators. These operators all have weight three. Operators of different types pairwise anti-commute, but they commute with all remaining stabilizers in the lattice.}\label{fig:breaking}
	\end{figure}

	Take a torus which is subdivided into hexagons.
	The quantum stabiliser code associated with this lattice is the hexagonal toric code, constructed by placing a data qubit on each edge of the lattice and associating each face and vertex with a $Z$ and $X$ stabiliser, respectively.
	Now consider a face~$f$ of the lattice and two of its vertices~$v$ and~$u$ that are not direct neighbours and do not have any neighbours in common either.
	We identify the vertices $u$ and $v$ and call the new vertex~$w$.
	This deforms the face~$f$ into a shape like a bow-tie with~$w$ in the center (see \Cref{fig:breaking}).
	Any edge which was incident to either~$v$ or~$u$ before is incident to~$w$ after this identification.
	
\begin{figure}[h]
		\centering
		\includegraphics{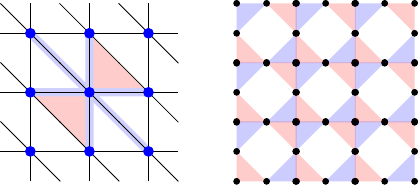}
		\caption{Left: Merging top-left and bottom-right vertices of all faces of a hexagonal tessellation leads to a tiling of bow-ties. The $X$- and $Z$-stabilizer belonging to the merged vertex in the center are highlighted in blue and red. Both are weight-6 operators.
		Right:~We can redraw the lattice by exchanging edges with vertices, representing the broken stabilizers as triangles.}\label{fig:toric_subsystem_code}
	\end{figure}

	There is a canonical subdivision of a bow-tie shaped face: we can simply consider either half.
	Similarly, the neighbourhood of a merged vertex in the middle of the bow-tie can be subdivided into two disjoint sets.
	In terms of the associated quantum code, there is a canonical way to {\em break} the $X$- and $Z$-checks associated with the vertices and faces (see \Cref{fig:breaking}, right).
	The four operators obtained by the breaking procedure are gauge operators and do not commute.
	Importantly, the gauge operators of a bow-tie while not commuting among themselves {\em commute with all other check operators}.
	By merging the upper left and lower right vertices of each hexagon, as shown in \Cref{fig:breaking}, we obtain the subsystem toric code of Ref.~\cite{bravyi2013subsystem}, shown as a tiling by bow-ties in \Cref{fig:toric_subsystem_code}~(left).

\begin{figure}
	\centering
	\includegraphics{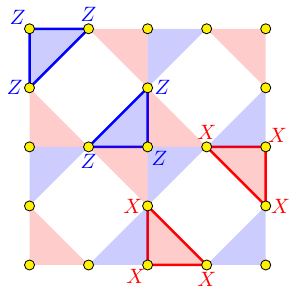}
	\caption{The subsystem toric code of Ref.~\cite{bravyi2013subsystem}. Data qubits (yellow filled circles) are placed in the middle of each edge and on each vertex of a square lattice of the toric code. Opposite sides are identified. The gauge group is generated by three-qubit \textit{triangle operators}. The two $Z$ triangle operators in the top left face are outlined with a blue border, and their product forms a 6-qubit $Z$ stabiliser. Similarly, in the bottom right face, two $X$ triangle operators are outlined with a red border, and their product is a 6-qubit $X$ stabiliser.}
	\label{fig:subsystem_toric_diagram}
\end{figure}

An alternative representation of the subsystem toric code can be found by placing a qubit on the middle of each edge and on each vertex of a square tiling. Each gauge operator is now represented by a triangle, with a qubit associated with each of its vertices. This lattice can be obtained from the bow-tie lattice by substituting edges for vertices. Borrowing terminology from \cite{bravyi2013subsystem}, each gauge generator is referred to as a \textit{triangle operator}, and consists of a Pauli operator acting nontrivially on its three qubits. There are four types of triangle operator in each face of the square lattice: two $Z$-type triangle operators defined in the north-west and south-east corners, and two $X$-type operators defined in the north-east and south-west corners. These four types of triangle operators are highlighted in \Cref{fig:subsystem_toric_diagram} for the $L=2$ subsystem toric code. Within each face, the product of each pair of $Z$-type triangle operators forms a 6-qubit $Z$ stabiliser, and the product of each pair of $X$-type triangle operators forms a 6-qubit $X$-type stabiliser. The subsystem toric code has $3L^2$ data qubits (there are $L^2$ vertices and $2L^2$ edges of the square lattice) and $2(L^2-1)$ independent stabiliser generators, forming a stabiliser code with $L^2+2$ logical qubits, $L^2$ of which are gauge qubits, with the remaining two logical qubits encoding quantum information. It can be verified that all triangle operators commute with the stabilisers and are therefore logical operators for the stabiliser code (since they are not stabilisers). The logical $\bar{Z}$ and $\bar{X}$ operators for each gauge qubit are chosen to be the north-west and north-east triangle operators of each face respectively. The remaining two pairs of logical operators are the same as for the toric code, each acting non-trivially only on data qubits lying on a (horizontal or vertical) homologically nontrivial loop of the torus. In \cite{bravyi2013subsystem} it was shown that the minimum distance of the subsystem toric code is $L$, and therefore the code has parameters $[[3L^2,2,L]]$.

In \cite{bravyi2013subsystem} a planar subsystem surface code was also introduced (with two qubit stabilisers on the boundary), which has code parameters $[[3L^2-2L,1,L]]$, and in \cite{brown2019handling} a planar rotated subsystem code was introduced (with three-qubit stabilisers on the boundary) which has parameters $[[\frac{3}{2}L^2-L+\frac{1}{2},1,L]]$. These compare to the parameters $[[2L^2,2,L]]$, $[[L^2+(L-1)^2,1,L]]$ and $[[L^2,1,L]]$ for the toric, planar and rotated surface codes respectively~\cite{dennis2002topological,bombin2007optimal}.

By mapping the threshold to the phase transition in the random-bond Ising model on the honeycomb lattice, the subsystem toric code has been found to have a threshold of around 7\% for maximum likelihood decoding, the independent $Z/X$ noise model and perfect syndrome measurements~\cite{bravyi2013subsystem}. Under the same noise model, the threshold using a minimum-weight perfect matching decoder is 6.5\%~\cite{fujii2012error}. Syndrome extraction can be done by measuring only the three-qubit triangle operators, and it has a threshold under a circuit-level depolarising noise model of around 0.6\%~\cite{bravyi2013subsystem}, which is below that of the standard surface code, which has a threshold approaching $1\%$ for a similar circuit-level depolarising noise~~\cite{raussendorf2007topological,wang2011surface,stephens2014fault}.

	\section{Finite-Rate LDPC Subsystem Codes}\label{sec:subsystem_hyperbolic_codes}
	
	\begin{figure}
	\includegraphics{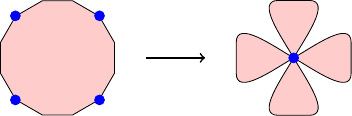}
		\caption{Merging vertices of a 12-gon. As this procedure effectively removes 3 independent $X$-checks we introduce 3 gauge qubits.}
		\label{fig:clover_merge}
	\end{figure}
	
	While the subsystem toric code can be derived from the hexagonal tessellation of a Euclidean surface, we will now show how we can also obtain subsystem codes derived from more general tessellations, including of hyperbolic surfaces. 
	A regular tessellation of a surface can be denoted by its Schl\"afli symbol $\{r,s\}$, which indicates that each face in the tessellation is an $r$-gon and $s$ faces meet at each vertex. 
	Regular tessellations of hyperbolic surfaces satisfy $1/r+1/s<1/2$.
	Hyperbolic codes, which are subspace codes derived from hyperbolic tessellations, have a finite encoding rate $k/n$ and distance scaling as $O(\log n)$~\cite{freedman2002z2,breuckmann2016constructions}, and it has been shown that they can require a smaller qubit overhead compared to the toric code and surface code for a target logical error rate under a phenomenological error model~\cite{breuckmann2017hyperbolic}.
	However, the stabiliser weight of hyperbolic codes is larger than for the toric code, making syndrome extraction more challenging, and a key benefit of the subsystem hyperbolic code construction we now give is that syndrome extraction can be done with only weight-3 check operators.
	Hyperbolic codes are a promising candidate for experimental realisation in systems that allow variable length connectivity between qubits~\cite{kollar2019hyperbolic}, such as modular architectures~\cite{nickerson2014freely,kalb2017entanglement}, and reduced check-weight simplifies stabiliser readout, as well as reducing crosstalk~\cite{chamberland2020topological}.
	
	A subsystem hyperbolic code can be obtained by merging multiple vertices of each face of a hyperbolic tessellation.
	 For example, a subsystem hyperbolic code can be constructed from a $\{12,3\}$ hyperbolic tessellation by merging four vertices of each 12-gon face, as shown for a single 12-gon in \Cref{fig:clover_merge}.
	In general, breaking an $m$-clover-shaped face introduces~$m$ local loop operators that do not mutually commute. 
	They will be interpreted as logical operators of the gauge qubits.	
	Note that for a clover with $m$-leaves we introduce $m-1$ linearly independent local loop operators as the product of all $m$ local loops is the original face, which is a stabilizer.

\begin{figure}
\includegraphics{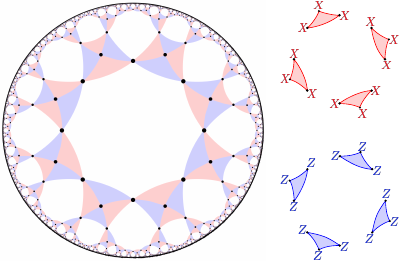}
\caption{The $\{8,4\}$ subsystem hyperbolic code. A qubit (each represented by a black filled circle) is placed in the center of each edge and on each vertex of an $\{8,4\}$ tessellation of a closed hyperbolic surface. A three qubit triangle operator is placed in each corner of each face. Each $X$ stabiliser is the product of the four $X$ triangle operators within a face (top right). Similarly, each $Z$ stabiliser is the product of the four $Z$ triangle operators within a face (bottom right).}
\label{fig:subsystem_hyperbolic_stabilisers}
\end{figure}

As for the subsystem toric and subsystem surface codes, there exists an alternative representation with qubits placed on vertices and where a triangle operator will be placed in each corner of each face of the hyperbolic tessellation. 
For example, we can construct an $\{8,4\}$ subsystem hyperbolic code by placing a triangle operators in the corner of each face of an $\{8,4\}$ tessellation, as shown in \Cref{fig:subsystem_hyperbolic_stabilisers}. Each $Z$ stabiliser is the product of all $Z$ triangle operators within a face of the $\{8,4\}$ tessellation, and similarly for $X$ stabilisers and $X$ triangle operators.
Note that this code obtained from placing triangle operators in an $\{8,4\}$ lattice can equivalently be constructed by merging faces in a $\{12,3\}$ tessellation.

We will adopt the former approach (qubits on vertices) and construct the subsystem hyperbolic codes directly by requiring that, as for the subsystem toric code, any pair of triangle operators that belong to the same face of the tessellation and overlap on a single qubit must be of opposite Pauli types. 
Similarly, any two triangle operators belonging to the same vertex of the original tessellation and overlapping on two qubits must be of the opposite Pauli type. 
In other words, adjacent triangle operators related by a single rotation about a face or a vertex must be of opposite Pauli types, and we will say that a tessellation that allows such an assignment of triangle operators is \textit{colourable}.
For a tessellation to be colourable, each face must have an even number of sides, and an even number of faces must meet at each vertex (so for regular $\{r,s\}$ tessellations, both $r$ and $s$ must be even).
Furthermore, to ensure that our stabilisers commute, we further require that four faces meet at each vertex of the tessellation.
In \Cref{app:subsystem_colouring} we show that a regular tessellation of a \textit{closed} surface is colourable if a particular function~$f$ (which we define) extends to a homomorphism from the symmetry group of the tessellation to the cyclic group~$\mathbb{Z}_2$.

\subsection{Properties of subsystem hyperbolic codes}

We will now consider some more properties of subsystem hyperbolic codes, each derived from a $\{2c, 4\}$ tessellation with edges $E$, vertices $V$ and faces $F$. Since we place a qubit on each vertex, and in the centre of each edge of this tessellation, our subsystem hyperbolic code will have $|E|+|V|$ data qubits. 
Each vertex in the tessellation has degree 4, and so $2|V|=|E|$. Furthermore, we also place~$n_a$ ancilla qubits within each triangle operator. While we can always use~$n_a=1$ ancillas per triangle operator by using schedules with some idle qubit locations (if necessary), we have parallelised many of our schedules which in some cases requires $n_a=2$. Each vertex is adjacent to four triangle operators and each triangle operator is adjacent to a single vertex. Therefore, in total there are $n=\frac{3}{2}|E|$ data qubits and $2n_a|E|$ ancilla qubits in our subsystem hyperbolic codes. For the subsystem toric code, where $|E|=2L^2$, there are~$3L^2$ data qubits and~$4n_aL^2$ ancilla qubits.

The number of faces in the $\{r,s\}$ tessellation satisfies $r|F|=2|E|$. Since the product of all $X$-type (or $Z$-type) stabilisers is the identity, and since these are the only relations the stabilisers satisfy, the number of independent stabilisers is $4|E|/r-2$. Therefore, the total number of logical qubits (including gauge qubits) is $(3/2-4/r)|E|+2$.

Aside from the triangle operators introduced within each face, the number of remaining bare logical operators (those in $C(\mathcal{G})\setminus \mathcal{G}$) is determined from the topology of the tessellation from which it is derived. Therefore, excluding gauge qubits, the number of logical qubits $k$ that a subsystem hyperbolic code derived from a $\{r,4\}$ tessellation encodes is given by~\cite{breuckmann2016constructions}
\begin{equation}
    k=\frac{|E|}{2}-\frac{2|E|}{r} + 2.
\end{equation}
This leaves $(1-2/r)|E|$ gauge qubits, or $r/2-1$ gauge qubits per face. The triangle operators act nontrivially on these gauge qubits.
The encoding rate of the subsystem hyperbolic code is therefore
\begin{equation}
    \frac{k}{n}=\frac{1}{3}-\frac{4}{3r}+\frac{2}{n}.
\end{equation}
There are $4n_a/3$ ancilla qubits per data qubit, leading to $(4n_a/3+1)n$ qubits in total. Note that this expression does not depend on $r$: the number of ancilla qubits is proportional to the number of data qubits, and the constant of proportionality is the same regardless of which $\{2c,4\}$ tessellation we use.

In \Cref{sec:subsystem_hyperbolic_distance}, we show that the distance $d$ of a subsystem hyperbolic or semi-hyperbolic code is bounded by $d_X/2\leq d \leq d_X$, where $d_X$ is the $X$ distance of the \textit{subspace} hyperbolic or semi-hyperbolic code derived from the same tessellation. The $X$ distance of the subspace code is always less than or equal to its $Z$ distance for the codes we consider, and so the distance of the subsystem code is at least half, and at most the same as, the distance of the subspace code. We analyse the distances of the codes we construct in \Cref{sec:subsystem_hyperbolic_distance}, and find codes with distances that span this full range.

\subsection{Condition for consistent scheduling}\label{sec:scheduling_condition}

\begin{figure}
	\centering
	\subfloat[]{
 	\includegraphics{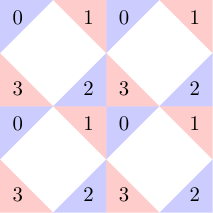}
	}
	~
	\subfloat[]{
	\includegraphics[width=0.45\columnwidth]{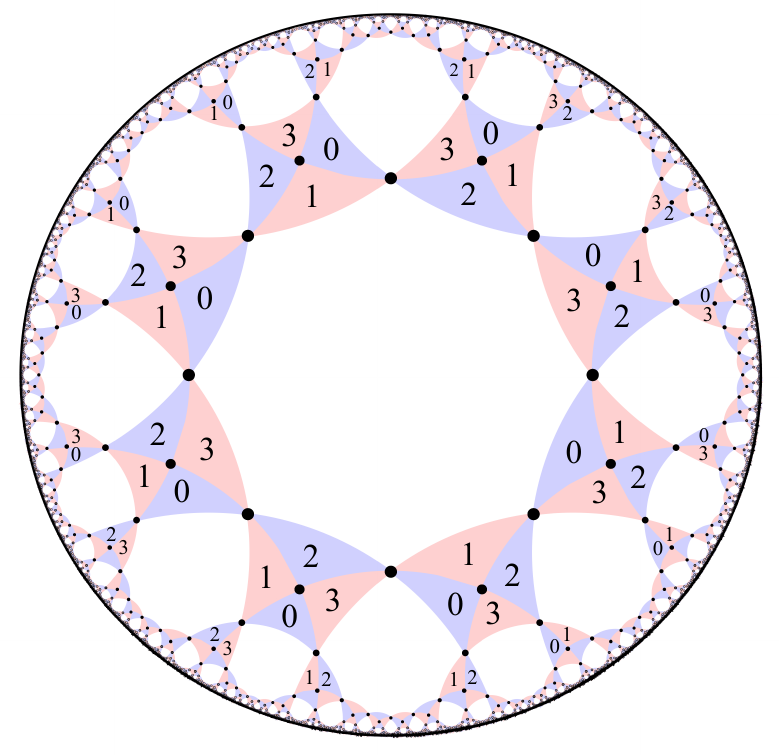}
	}
	\caption{(a) An $L=2$ subsystem surface code. The four types of triangle operators are labelled as 0, 1, 2 and 3. (b)~Labelling of the four types of triangle operators on an $\{8,4\}$-tessellation of the hyperbolic plane. The neighbourhood of each triangle operator (the types and relative locations of triangle operators it overlaps with) is the same as in the toric code.}\label{fig:triangle-labelling}
	\end{figure}

In order to determine the syndrome used for decoding, we require a \textit{stabiliser measurement schedule}, which is the sequence of gates applied to data and ancilla qubits in order to measure the eigenvalues of the stabilisers. We will now show that any valid stabiliser measurement schedule defined within a single face of the subsystem toric code and chosen to be periodic in space (i.e.~identical for every vertex or face) can be generalised for a subset of $\{4c,4\}$ subsystem hyperbolic codes, for $c\in\mathbb{Z}^+$. 
The measurement schedule used by \citeauthor{bravyi2013subsystem}~\cite{bravyi2013subsystem} is an example of such a periodic schedule.

We first assign an element of the cyclic group $\mathbb{Z}/4\mathbb{Z}$ to each of the four types of triangle operators within a face, and will call such an assignment a \textit{labelling}. 
We choose to label the north-west, north-east, south-east and south-west triangle operators with the elements 0, 1, 2 and 3 of $\mathbb{Z}/4\mathbb{Z}$, respectively (see \Cref{fig:triangle-labelling}(a)). Note that, for a translationally invariant schedule, each triangle operator with a given $\textit{label}$ in the subsystem toric code is assigned an identical schedule. Triangle operators with different labels have different measurement schedules. In order to apply this measurement schedule to the subsystem hyperbolic code, we label every triangle operator as one of these four types in such a way that the schedule always looks locally the same as for the subsystem toric code to ensure that it remains correct. More precisely, for each triangle operator with a given label in the subsystem hyperbolic code, its neighbourhood of triangle operators it shares qubits with (and their labels) must be the same as for a triangle operator with the same label in the subsystem toric code. We will call a labelling that achieves this a \textit{valid labelling}, and a \textit{schedulable} code is one that admits a valid labelling. In \Cref{app:labelling_homomorphism}, we show that a regular tessellation of a closed hyperbolic surface admits a valid labelling if a particular function $h$ (which we define) extends to a homomorphism from the proper symmetry group of the tessellation to the cyclic group $\mathbb{Z}/4\mathbb{Z}$. We show that a subset of $\{4c,4\}$ regular tessellations of closed hyperbolic surfaces satisfy this property. An example of a valid scheduling of the $\{8,4\}$ tessellation of the hyperbolic plane is shown in \Cref{fig:triangle-labelling}(b).

\subsection{Subsystem semi-hyperbolic codes}\label{sec:semi_hyperbolic_codes}

The $\{8,4\}$ subsystem hyperbolic code has stabilisers of weight 12, which is double that of the subsystem toric code. Despite the check operators still being weight 3, we find that the large stabiliser weight results in a lower threshold of $0.31(1)\%$ compared to $0.666(1)\%$ for the subsystem toric code. 
The intuition behind this is the following: if a stabiliser has higher weight, it provides less information about the location of an error and requires more gates to be used when measured, making it harder to measure precisely.

To address this issue, we can construct subsystem codes derived from \textit{semi-hyperbolic} tilings, introduced in Ref.~\cite{breuckmann2017hyperbolic}.
The idea is to fine-grain the tessellation leading to lower-weight stabilizers. 
A semi-hyperbolic tiling is derived from a $\{4,q\}$ regular tessellation of a closed hyperbolic manifold for $q>4, q\in \mathbb{Z}^+$. Each (square) face of the $\{4,q\}$ tessellation is tiled with an $l\times l$ square lattice. By doing so, the curvature of the surface is weakened. The subspace quantum code derived from the semi-hyperbolic tessellation (a semi-hyperbolic code) has larger distance and reduced check weight compared to a code derived from the original $\{4,q\}$ tessellation. This comes at the cost of requiring $l^2$ times more qubits and, since the number of logical operators is unchanged, the encoding rate is reduced by a factor of $l^2$. An important advantage of semi-hyperbolic codes is that, by increasing $l$, we obtain a family of codes with distance scaling like $\sqrt{n}$ (as for the toric code), while expecting to retain a reduced qubit overhead relative to the toric code~\cite{breuckmann2017hyperbolic}. The same advantages apply for the subsystem semi-hyperbolic codes we construct in this work.

Recall that the tessellations that we derive subsystem hyperbolic codes from must have vertices of degree four, and each face must have $4c$ sides (where $c\in\mathbb{Z}^+$). On the other hand, a $\{4,q\}$ semi-hyperbolic tiling instead has faces with four sides, while vertices have degree four or $q$. We can therefore derive a subsystem code from the \textit{dual lattice} of $\{4,4c\}$ semi-hyperbolic tessellation. In \Cref{app:labelling_homomorphism} we show that if an $\{8,4\}$ tessellation is schedulable, then so is the semi-hyperbolic tessellation derived from it. Therefore, each schedulable closed $\{8,4\}$ tessellation defines a family of subsystem semi-hyperbolic codes (each code in the family having a different lattice parameter $l$), and where each code in the family is schedulable.

We say that an $l$, $\{4c,4\}$ subsystem semi-hyperbolic code is the code derived by placing a triangle operator in each corner of each face of the dual lattice of a semi-hyperbolic lattice, where that semi-hyperbolic lattice was constructed by tessellating each face of the $\{4,4c\}$ tessellation with an $l\times l$ square lattice. The subsystem semi-hyperbolic codes we construct and analyse in this work are $l=2$, $\{8,4\}$ subsystem semi-hyperbolic codes. The irregular tessellations these codes are derived from therefore contain both square and octagonal faces, with four faces meeting at each vertex.

\section{Improved Error Correction by Gauge-Fixing}\label{sec:gauge_fixing}

We will now introduce some general techniques that improve the quantum error correcting performance of a wide class of subsystem codes. We will alter the stabiliser measurement procedure \textit{in software}, in such a way that the individual gauge operator measurements themselves yield useful information. This is in contrast to existing methods for decoding subsystem codes in the literature, where individual gauge operator measurements themselves are never treated as syndrome bits, and only their products (the stabilisers) are used for decoding.

	While we will analyse these techniques numerically using the subsystem code constructions given in \Cref{sec:2d_construction} and \Cref{sec:subsystem_hyperbolic_codes}, the key ideas can be applied to the vast majority of subsystem codes considered in the literature, for which stabiliser eigenvalues can be inferred by measuring gauge operators.
	In fact, these techniques address one of the main drawbacks of subsystem codes, which is that they typically have lower thresholds.
	Low thresholds arise partly because stabiliser eigenvalues are determined by combining the outcomes of many gauge operator measurements, each of which may be faulty, making their measurement less reliable.
	Additionally, these high weight stabilisers provide less information about which qubit has suffered an error, further reducing the threshold.
	The most dramatic example of this effect is the Bacon-Shor code which, although it has weight-2 check operators, has no threshold, as the stabilizer operators grow with system size.
	The techniques we introduce can also be used when applying logical operations with \textit{subspace} codes, as we explain in \Cref{sec:lattice_surgery}, since lattice surgery and code deformation for surface codes can be interpreted as gauge fixing of a larger subsystem code~\cite{vuillot2019code}.

	We call the general method \textit{schedule-induced gauge fixing}, since we will be altering the schedule of the stabiliser measurement circuits in such a way that gauge fixing can be used to significantly improve the error correcting performance when decoding. We will refer to it simply as gauge fixing when the meaning is clear from context.

	Schedule-induced gauge fixing can be applied to a large class of subsystem codes, for which there are stabilisers $s$ that are the product of gauge operators, $s=g_0g_1\ldots g_{m-1}, \quad g_i\in \mathcal{G}\setminus \mathcal{S}$. 
	We call these gauge operators \textit{gauge factors} $\mathcal{G}^s$ of $s$,
	\begin{equation}
	\mathcal{G}^s:=\{g_0,\ldots,g_{m-1}|g_i\in\mathcal{G}\setminus\mathcal{S}, s=g_0g_1\ldots g_{m-1}\},
	\end{equation}
and stabilisers which admit such a decomposition will be referred to as \textit{composite stabilisers}. In general there can be more than one such decomposition for a given stabiliser, though we are typically most interested in the \textit{minimum-weight decomposition}, where the average weight of gauge factors $g_i\in\mathcal{G}\setminus\mathcal{S}$ is minimised. For the codes we construct in this work there is a unique minimum-weight decomposition for each stabiliser, though in general there can be more than one~\cite{brown2016fault}. For CSS subsystem stabiliser codes the gauge factors of each stabiliser mutually commute, and can be measured in any relative order. For more general subsystem codes, the order of measurements of gauge factors $g_0g_1\ldots g_{m-1}$ of each stabiliser $s\in\mathcal{S}$ must be chosen such that each gauge factor measurement $g_i$ commutes with the product $g_0g_1\ldots g_{i-1}$ of gauge factor measurements before it. In Ref.~\cite{suchara2011constructions}, this condition was shown to be both necessary and sufficient to guarantee that the stabiliser can indeed be recovered from the product of individual measurements.
Schedule-induced gauge fixing will typically be most useful for subsystem codes which have at least one composite stabiliser, and for which the weight of each composite stabiliser is greater than the weight of each of its gauge factors. In the case of the subsystem codes studied in this work, the gauge factors of each $Z$ stabiliser associated with a face are the $Z$ triangle operators belonging to that face (and similarly for $X$ stabilisers and $X$ triangle operators). 

When decoding subsystem codes with existing methods, the syndrome used consists of eigenvalues of stabilisers. In other words, where a stabiliser is composite, measured by taking the product of the measurements of its $m$ gauge factors $g_i\in \mathcal{G}^s$, it is the product that is used, not the result of each gauge factor measurement individually. Therefore, for each stabiliser, we are measuring $m$ bits of information, and only using a single bit (their parity) for decoding. For the most simple stabiliser measurement schedules typically used, the parity is indeed all the useful information that can be used for decoding. This is because $\mathcal{G}$ is not abelian and, by definition, each gauge factor $g_i\in \mathcal{G}^s$ must anti-commute with at least one other gauge operator $h\in\mathcal{G}$. Once $h$ is measured, either $h$ or $-h$ becomes a stabiliser, and a subsequent measurement of $g_i$ will result in either $1$ or $-1$ at random with $P(1)=P(-1)=0.5$. Consider a schedule~$W$ of measurements of check operators $K_0K_1\ldots K_{N-1}$, chronological order from left to right, where each check operator $K_i$ is either a gauge factor or a stabiliser that is not composite, and where each $K_i$ is measured once. If this measurement schedule~$W$ is simply repeated periodically, then every consecutive pair of measurements of any check operator~$K_i$ will be separated by one measurement of every other check operator. As a result, if the check operators in~$W$ generate $\mathcal{S}$ as required, every measurement of a gauge factor will give a random outcome and will not be useful for decoding, since its eigenvalue will not have been preserved between consecutive measurements. In fact, the eigenvalue of any product of check operators that is not in $\mathcal{S}$ will also not be preserved between consecutive measurements, following similar reasoning.

However, we can instead choose a measurement schedule $W$, again repeated periodically, where some gauge factors $g_i$ are measured multiple times within $W$, with no anti-commuting check operators measured between consecutive measurements of $g_i$ within $W$. In this case, only the first measurement of $g_i$ in $W$ will have a uniform random outcome, whereas the remaining measurements of $g_i$ within $W$ will have fixed outcomes (if no error as occurred), as the quantum state will (temporarily) be an eigenstate of $g_i$---we can think of $g_i$ (or $-g_i$) as a temporary stabiliser.

\subsection{Gauge-fixing for CSS codes}

We will now restrict our attention to CSS subsystem codes, for which the gauge group $\mathcal{G}$ can be decomposed into a set of operators each in $\{I,X\}^n$, which we denote $\mathcal{G}_X$, and a set of operators each in $\{I,Z\}^n$, which we denote $\mathcal{G}_Z$, with $\mathcal{G}=\mathcal{G}_X\cup\mathcal{G}_Z$. The stabiliser group can similarly be decomposed into either $X$-type or $Z$-type Pauli operators. For CSS subsystem codes, the most common measurement schedule consists of alternating between measuring all $X$-type and all $Z$-type check operator measurements in a repeating sequence. In other words, the sequence of measurements for measuring the $X$ or $Z$ stabilisers is of the form $(ZX)^r$, where $2r$ is the number of rounds of stabiliser measurements, and the chronological order is from left to right. We call such a sequence of measurements a \textit{homogeneous} schedule, since all stabilisers of the same Pauli-type are given identical measurement schedules. Equivalently, for the subsystem codes we construct, a homogeneous schedule assigns the same schedule to each face of the lattice from which it is derived. We will sometimes denote a schedule just by its longest repeating subsequence if the number of repetitions is not relevant (i.e.~denoting the above schedule by $ZX$ rather than $(ZX)^r$). 

For the $ZX$ schedule, each $X$ gauge operator measurement comes directly after the measurement of a $Z$ gauge operator that it anti-commutes with (and vice versa), and so the outcome of each individual gauge operator measurement is random. 
However, by repeating $X$ or $Z$ check operator measurements we can temporarily \textit{fix} some gauge operators as stabilisers. 
As an example, consider a homogeneous schedule of the form $(Z^2X^2)^r$. The first in each pair of $X$ gauge operator measurements will give a random outcome, whereas the second is simply a repetition of the first and, provided no error has occurred, will give the same outcome as the first measurement.
	The same is true for the first and second $Z$ gauge operator measurement outcomes.
	
	\subsection{Gauge fixing matching graph: vertex splitting and merging}\label{sec:gauge_fixing_matching}
	
	We now show how this additional gauge operator information can be used when decoding a CSS subsystem code using a method based on minimum-weight perfect matching, which introduces the additional requirement that the code must have no more than two stabilisers of a given Pauli type acting non-trivially on each qubit. Subsystem codes which satisfy these properties include the subsystem surface code~\cite{bravyi2013subsystem}, the Bacon-Shor code~\cite{bacon2006operator,aliferis2007subsystem}, and some 2D compass codes~\cite{li20192d}, including heavy-hexagon codes~\cite{chamberland2020topological}.
	
	\begin{figure}
\includegraphics[width=0.35\columnwidth]{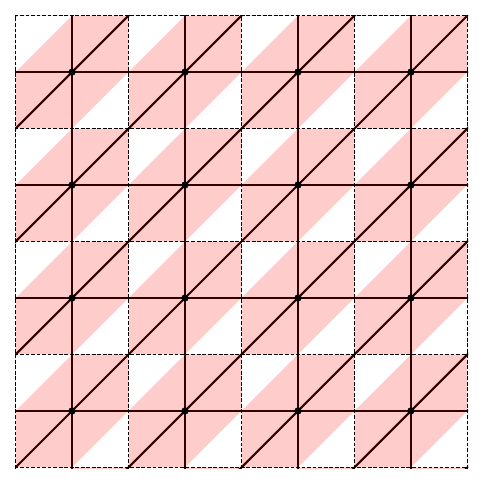}
\includegraphics[width=0.35\columnwidth]{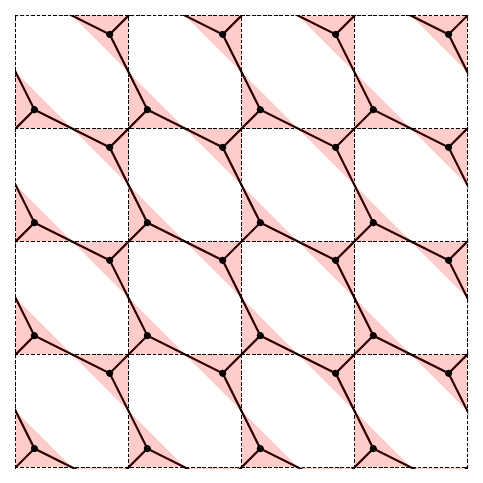}
\caption{Matching graphs ($X$-type) for the subsystem toric code with no triangle operators fixed as stabilisers (left) and all triangle operators fixed as stabilisers (right).}
\label{fig:hexagonal_and_triangular_matching_graphs}
\end{figure}

	As an example, let us first consider the 2D matching graphs of the subsystem toric code, assuming perfect stabiliser measurements. Each vertex in the $X$-type (or $Z$-type) matching graph corresponds to an $X$ (or $Z$) stabiliser, and each edge corresponds to a qubit (and therefore a possible error). For the stabiliser group of the subsystem toric code with no gauge operators fixed, both the $X$-type and $Z$-type matching graphs are triangular lattices, as shown in \Cref{fig:hexagonal_and_triangular_matching_graphs} (left) for the $X$-type matching graph. This triangular lattice matching graph has a minimum-weight perfect matching (MWPM) threshold of 6.5\% with perfect measurements~\cite{fujii2012error}. However, once we have measured all the $X$-type gauge operators, they become gauge-fixed as stabilisers (up to signs that can be corrected in software), and the stabiliser group we obtain is that of the hexagonal toric code~\cite{fujii2012error}. The new associated $X$-type matching graph instead has an improved MWPM threshold with perfect measurements of 15.6\%~\cite{fujii2012error}, exceeding that of the toric code on a square lattice of 10.3\%~\cite{dennis2002topological}. If we measure all the $Z$-type gauge operators, we instead obtain the dual of the hexagonal toric code, and now the $Z$-type matching graph is a hexagonal lattice.
	
	When using the standard $ZX$ schedule for the subsystem toric code, the stabiliser group is indeed constantly switching (up to signs) between the hexagonal toric code and its dual, both abelian subgroups of the gauge group~$\mathcal{G}$. However, each gauge operator is only ever fixed immediately after it is measured, and is randomised by the time the same gauge operator is next measured, since an anti-commuting gauge operator of the opposite Pauli-type is measured in between these consecutive measurements of the same gauge operator. However, by making more than one consecutive measurement of gauge operators of a given Pauli type, we will now show that we can gauge fix into the hexagonal toric code (and its dual) for longer durations, thereby making more valuable use of the individual gauge operator outcomes themselves.
	
	\begin{figure}
	\includegraphics{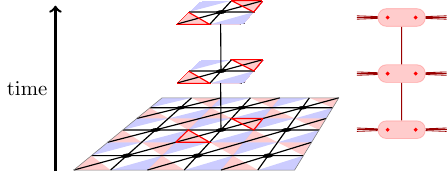}
	\caption{The 3D $X$-type matching graph for the subsystem toric code. Left: We show three time steps of the 3D matching graph for a single $X$ stabiliser (highlighted in red), with black lines denoting edges. Right: We also use this more simple, abstract notation to depict the same 3D matching graph in our work, restricted to a single face of the lattice. Here, each pale red rounded rectangle corresponds to an $X$ stabiliser in one of three consecutive time steps in the matching graph. Red dots denote $X$ triangle operator measurements (two of which within a face form a stabiliser), and red lines denote edges in the 3D matching graph.}
	\label{fig:matching_3d_layers}
	\end{figure}
	
	Since measurements themselves can be faulty, we must instead use a 3D matching graph when decoding the subsystem toric code. Each vertex in the matching graph corresponds to a stabiliser measurement, and each edge $(u,v)$ corresponds to a single fault that can occur, creating a defect (-1 syndrome) at vertices $u$ and $v$. In order to handle measurement errors, each stabiliser measurement is repeated $T\geq L$ times~\cite{dennis2002topological}, and a syndrome for a stabiliser at time step $t$ takes the value $-1$ if its value differs from its measurement in time step $t-1$. Measurement errors correspond to time-like edges, and memory (data qubit) errors correspond to space-like edges. There are also single circuit faults that can induce \textit{diagonal} edges, which have vertices that differ in both space and time. We can label each vertex in the matching graph with a coordinate $(s,t)$, where $t$ is the time step and $s=g_0\ldots g_{m-1}$ denotes the stabiliser using its gauge factors $g_i\in\mathcal{G}^s$. We depict the 3D matching graph for the subsystem toric code in \Cref{fig:matching_3d_layers}.
	
\begin{figure}
	\centering
	\includegraphics{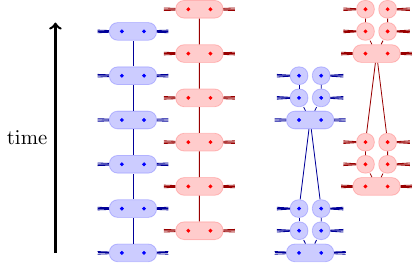}
	\caption{Matching graph for a single face of the subsystem toric code using a homogeneous $(ZX)^6$ schedule (left) and a homogeneous $(Z^3X^3)^2$ schedule (right). The vertical axis corresponds to time, with the direction of time being from bottom to top. Small blue and red filled circles correspond to $Z$ and $X$ gauge operator measurements respectively, with each vertical column of small filled circles corresponding to a single gauge operator. Large light blue and light red filled rounded rectangles (or rounded squares) correspond to stabilisers, being the product of the gauge operators they enclose. Diagonal edges (between stabilisers that differ in space and time) have been omitted for clarity. Blue and red lines correspond to edges in the $Z$ and $X$ matching graphs, respectively.}\label{fig:matching_graph}
\end{figure}

	For the $ZX$ schedule used in the previous literature, gauge operators are never fixed and stabilisers are always the product of gauge operators, whereas for many of the schedules we use, we can fix a subset of the gauge operator measurements, and obtain (temporarily) stabilisers consisting of single gauge operators. In our matching graph, we can fix a measurement of a gauge operator $g$ as a stabiliser if no gauge operator $h$ which anti-commutes with $g$ has been measured since the last measurement of $g$. This is demonstrated in \Cref{fig:matching_graph} for the schedules $(ZX)^6$ and $(Z^3X^3)^2$. For the $(ZX)^6$ schedule, gauge operators can never be fixed as stabilisers in the matching graph, whereas for the $(Z^3X^3)^2$ schedule, two-thirds of the gauge operator measurements can be fixed as stabiliser measurements. Since each gauge operator has weight 3, by fixing some gauge operators as stabilisers, we can reduce the weight of some stabiliser measurements from~6 down to~3.
	
Since the stabilisers can change between consecutive time steps when using gauge fixing, we must generalise our definitions of the difference syndrome and vertical edges in the matching graph. For our generalised difference syndrome, we set the syndrome of stabiliser $s$ to be $-1$ in time step $t$ if its eigenvalue differs from that of the same product of gauge operators in time step $t-1$. We draw a vertical edge in the matching graph between a stabiliser measurement $s_t$ in time step $t$ and measurement $s_{t-1}$ in time step $t-1$ if $s_t$ and $s_{t-1}$ have at least one gauge factor in common. 

As an example we will now consider the case where a stabiliser has two gauge factors, as is the case for the subsystem toric code. Suppose a stabiliser is the product $g_0g_1$ of gauge factors $g_0$ and $g_1$ in time step $t-1$, but both $g_0$ and $g_1$ are fixed as stabilisers in time step $t$. We say that the stabiliser vertex is \textit{split} into two vertices in time step $t$, with the matching graph locally looking like (with time propagating upwards):
\begin{center}
\begin{tikzpicture}
\pairstab{0}{\zlcol}{\zcol}{\bl}{\br}{bs}{0};
\singlestab{1}{\zlcol}{\zcol}{\bl}{\br}{bsl}{bsr};
\draw[\zlcol] (bs0) -- (bsl1);
\draw[\zlcol] (bs0) -- (bsr1);
\end{tikzpicture}
\end{center}
and a measurement error in time step $t-1$ on gauge factor measurement $g_0$, e.g.~at the vertex $(g_0g_1,t-1)$, will cause a $-1$ difference syndrome at vertex $(g_0g_1,t-1)$ as well as vertex $(g_0,t)$. Therefore, this measurement error corresponds to flipping the vertical edge $((g_0g_1,t-1), (g_0,t))$. The same argument holds for a measurement error on~$g_1$ in time step~$t-1$ corresponding to flipping the other vertical edge $((g_0g_1,t-1), (g_1,t))$. Similarly, we can fix~$g_0$ and~$g_1$ as stabilisers in time step $t-1$ but instead have the stabiliser~$g_0g_1$ in time step~$t$ (the vertices are \textit{merged} in time step $t$). This would be the case if gauge operators are measured in between time steps~$t-1$ and~$t$ that anti-commute with~$g_0$ and~$g_1$.
The matching graph locally looks like:
\begin{center}
\begin{tikzpicture}
\pairstab{1}{\zlcol}{\zcol}{\bl}{\br}{bs}{0};
\singlestab{0}{\zlcol}{\zcol}{\bl}{\br}{bsl}{bsr};
\draw[\zlcol] (bs1) -- (bsl0);
\draw[\zlcol] (bs1) -- (bsr0);
\end{tikzpicture}
\end{center}
and we find that a measurement error that occurs at the vertex $(g_0,t-1)$ results in a $-1$ syndrome at both $(g_0,t-1)$ and $(g_0g_1,t)$, corresponding to flipping the edge $((g_0,t-1),(g_0g_1,t))$. Similarly, a measurement error at vertex $(g_1,t-1)$ corresponds to flipping the edge $((g_1,t-1),(g_0g_1,t))$. While, in this example, we have considered stabilisers which have only two gauge factors (which is the case for subsystem toric codes), the definition of the difference syndrome can be applied to stabilisers with any number $m$ of gauge factors. For example, we have $m=4$ for the $\{8,4\}$ subsystem hyperbolic codes considered in this work, since these have four triangle operators (gauge factors) in each face.

In a stabiliser round in which all gauge operators are fixed (matching graph vertices are split), there are two distinct advantages which gauge fixing can offer. Firstly, vertical time-like edges have a lower error probability, since the syndrome corresponding to a vertex is obtained from only a single check operator measurement, rather than taking the product of multiple measurements. Secondly, the degree of vertices in the matching graph is reduced. 

The advantage that this can offer becomes clear when we again consider the (space-like) matching graph of the subsystem surface code when all gauge operators are fixed, compared to the matching graph when they are not fixed. We have found that the hexagonal lattice matching graph when gauge operators are fixed (\Cref{fig:hexagonal_and_triangular_matching_graphs}, right) has a threshold of around $4.1\%$ under a phenomenological noise model. On the other hand, for the triangular lattice matching graph when no gauge operators are fixed (\Cref{fig:hexagonal_and_triangular_matching_graphs}, left) we find a threshold of 2.0\% with a phenomenological noise model (see \Cref{fig:phenomenological_triangular}). Furthermore, the outcomes of the weight-three checks are more reliable, since their measurement circuits are shorter. However, a potential disadvantage of gauge fixing is that by repeating $X$ checks, more errors accumulate for the next measurement of $Z$ checks, for which $Z$ gauge operators cannot be fixed. 
We will show in \Cref{sec:gauge_fixing_depol} how this trade-off leads to an optimal homogeneous schedule for the threshold under a circuit-level depolarising noise model.

\subsection{Homogeneous stabiliser measurement circuits}

In order to measure the triangle operators (and therefore stabilisers), we require a circuit to measure each triangle operator using an ancilla qubit.
We will now show how these circuits can be constructed for \textit{homogeneous schedules}, where the same schedule is applied to each face in the lattice.
As discussed in \Cref{sec:scheduling_condition}, all triangle operators with the same label share the same schedule in the subsystem toric, hyperbolic and semi-hyperbolic codes we use, so we need only specify four parity check circuits, one for each label. Each triangle operator consists of three data qubits and at least one ancilla, and can be measured using three CNOT gates, along with state preparation and measurement of an ancilla. A time step is defined as the time taken for a CNOT gate, and we assume that state preparation and measurement combined take a single time step. This is similar to the assumption of non-demolition measurements in Refs.~\cite{wang2011surface,fowler2012towards}, except we will assume both state preparation and measurement errors, rather than just the latter. In Ref.~\cite{bravyi2013subsystem} the authors instead assume that state preparation and measurement each take a time step, and use an additional ancilla to parallelise state preparation and measurement into a single time step. The parity check measurement circuit therefore takes four time steps.

\begin{figure}
\centering
\includegraphics{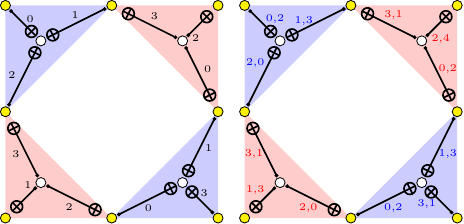}
\caption{Parity check measurement schedule for the subsystem surface code using a homogeneous $(ZX)^r$ sequence (left)~\cite{bravyi2013subsystem}, a homogeneous $Z^r$ sequence (right, blue text) and a homogeneous $X^r$ sequence (right, red text). CNOT gates are labelled with the time step(s) they are applied in, which are given modulo 4, since all schedules have period 4.}\label{fig:schedule}
\end{figure}

The measurement schedules we use are shown in \Cref{fig:schedule}. The schedule shown on the left of \Cref{fig:schedule} is for alternating measurement of the Pauli-$Z$ and Pauli-$X$ operators ($ZX$ schedule), and is the same as that used in Ref.~\cite{bravyi2013subsystem}. The right hand diagram in \Cref{fig:schedule} shows the schedule for measuring $ZZ$ (blue labels) as well as the schedule for measuring $XX$ (red labels). All three of these schedules have period 4, and so the time steps which each gate is labelled with are given modulo 4. Note that the first half of the $ZZ$ schedule matches the $Z$ component of the $ZX$ schedule, and the first half of the $XX$ schedule matches the $X$ component of the $ZX$ schedule. Therefore, the schedule for \textit{any} homogeneous sequence can be implemented by concatenating these three schedules (or subsets of them). For the standard $ZX$ schedule, we need only a single ancilla qubit for each triangle operator. For schedules which contain $ZZ$, we use two ancillas per $Z$ triangle operator to parallelise consecutive triangle operator measurements, and similarly we use two ancillas per $X$ triangle operator for parallelised schedules containing $XX$.

For the subsystem hyperbolic and semi-hyperbolic codes, we generalise the schedule in \Cref{fig:schedule} by using the same schedule for triangles with the same label, as explained in \Cref{sec:scheduling_condition} and \ref{app:labelling_homomorphism}. Each individual fault in the measurement circuit results in at most a single data qubit error, a property that is made possible by the weight-three gauge operators. As a result of this bare-ancilla fault tolerance of the measurement circuits, we can correct up to the full code distance for all the codes we have constructed.

\subsection{Edge weights}

In order to decode the subsystem surface codes using minimum-weight perfect matching, we construct a matching graph, where each individual fault that can occur flips an edge in the matching graph~\cite{dennis2002topological, bravyi2013subsystem}. We assign each edge a weight $w=\log((1-p)/p)$, where $p$ is the total probability that any individual fault will result in the edge being flipped~\cite{dennis2002topological,wang2011surface,huang2020fault}. 

\begin{figure}
\includegraphics{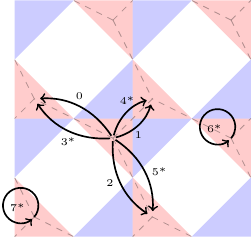}
\caption{The different types of edges in the 3D matching graph of the subsystem surface code for $X$-type checks only, when all $X$-type gauge operators are fixed. Each unique edge type is labelled with a number. If an asterisk is present in the label, the edge is from time step $t$ to $t+1$, otherwise the edge is purely space-like. The whole $X$ matching graph for a single time step is drawn with grey dashed lines.}\label{fig:matching_graph_edges}
\end{figure}

We will first consider the matching graph obtained by only measuring $X$-type check operators and fixing \textit{all} $X$-type gauge operators as stabilisers. We will see later that all other matching graphs for arbitrary homogeneous schedules can be obtained by merging edges and/or vertices in this matching graph. There are two types of $X$-type gauge operators in the subsystem surface code, as shown in \Cref{fig:triangle-labelling}, labelled by 1 and 3, which we will refer to as $T_1$ and $T_3$, respectively. Every space-like or diagonal edge is from a $T_1$ to a $T_3$ (or vice versa), and the neighbourhood of every triangle operator with the same label is identical. All seven types of edges in the matching graph for $X$-type checks are shown in \Cref{fig:matching_graph_edges}. All edges are undirected, but are denoted by directed arrows in the diagram to remove any ambiguity in the definition of the diagonal edges. The purely space-like edges are labelled 0, 1 and 2, purely time-like errors are labelled 6 and 7 and diagonal edges are labelled 3, 4 and 5. Diagonal and time-like errors are drawn from time step $t$ to time step $t+1$, whereas space-like edges connect vertices within a single time step. 
Therefore, each vertex in this matching graph has degree~8 (since each vertex is both the source and target of a time-like edge).

If an $X$-type check operators is measured directly after a $Z$-type check operator that anti-commutes with it, then this $X$-type check operator cannot be fixed, and the matching graph shown in \Cref{fig:matching_graph_edges} is not quite valid. However, we can use the vertex merging procedure detailed in Section~\ref{sec:gauge_fixing_matching} to give the matching graph the correct structure. When the $X$-type check operators within a face of the lattice cannot be fixed, then the corresponding $X$-type matching graph vertices from that face (each vertex $v_{g_i}$ corresponding to a gauge factor~$g_i$) are \textit{merged} into a single vertex $v_s$. The edges incident to $v_s$ each correspond to an edge incident to a gauge factor vertex~$v_{g_i}$. This process can result in more than one edge (a multi-edge) between the same pair of vertices (such as for time-like edges in homogeneous $(ZX)^r$ schedules). When this happens, we replace the multi-edge with a single edge, and assign it a flip probability equal to the probability that an odd number of edges in the multi-edge would flip.

\begin{table}
\begin{tabular}{cccccc}
Edge type   & $G_1^X$ & $G_1^Z$ & $G_2^Z$ & $P_X$ & $M_X$ \\
\hline
0 & 2 &$2r_Z$&0&0&0 \\
1 & 2 &$2r_Z$&$2r_Z$&0&0 \\
2 & 2 &$2r_Z$&$2r_Z$&0&0 \\
3 & 2 &0&0&0&0 \\
4 & 2 &0&0&0&0 \\
5 & 2 &0&0&0&0 \\
6 & 3 &0&0&1&1 \\
7 & 3 &0&0&1&1 \\
\end{tabular}
\caption{Number of single faults that can cause each type of edge to flip in the 3D matching graph for $X$-type check operators. Each $G_1^X$ or $G_1^Z$ fault is a single Pauli error arising from a $\mathrm{CNOT}$ gate in the measurement circuit for an $X$-type or $Z$-type gauge operator respectively. Each $G_2^Z$ fault is a \textit{pair} of Pauli errors arising from a single $\mathrm{CNOT}$ gate in the measurement circuit for a $Z$-type gauge operator. $P_X$ and $M_X$ are state preparation and measurement errors in the $X$-type check operator measurement schedule, respectively. $r_Z$ is the number of rounds of $Z$-type check operator measurements that have occurred since the last $X$-type check operator measurement. For example,~$r_Z=1$ always for $(ZX)^r$ schedules, and~$r_Z=2$ always for~$(ZZX)^r$ schedules. The edge types are shown in \Cref{fig:matching_graph_edges}. Faults for the $Z$ matching graph can be found by exchanging $Z$ and $X$ in the table.}\label{table:edge_faults}
\end{table}

In order to calculate the probability $p$ that each edge flips (both for edge weights and for simulations), we count the number of single faults (of each type) that can lead to each type of edge flipping. These are given in Table~\ref{table:edge_faults} for the~$X$ matching graph (for $X$-type check operators).
The operators~$G_1^X$ and~$G_1^Z$ are Pauli errors from $\mathrm{CNOT}$ gates in the~$X$ or~$Z$ measurement schedule respectively, corresponding to either a~$XI$,~$IX$ or~$XX$ error acting after the gate. In the standard depolarising model,~$G_1^X$ or~$G_1^Z$ errors occur with probability $4p/15$. See Table~\ref{table:error_model} for the gate error probabilities under the independent noise model we use. $G_2^Z$ errors correspond to a pair of~$G_1^Z$ errors from the same $\mathrm{CNOT}$ gate in the $Z$ measurement circuit that both cause the same edge to flip. For example, both~$XI$ and $XX$ errors on a $\mathrm{CNOT}$ gate may cause the same edge to flip, and since these errors are mutually exclusive on the same gate, the chance of either of these errors occurring is exactly twice the probability that one of them occurs. The number of~$G_1^Z$ or~$G_2^Z$ errors that can cause an edge to flip depends on~$r_Z$, the number of~$Z$ check operator measurements that have occurred since the most recent prior~$X$ check operator measurement. We can recover the matching graph for the standard $(ZX)^r$ schedule used in Ref.~\cite{bravyi2013subsystem} by setting~$r_Z=1$ and merging all vertices within each face (up to small differences in the error model, shown in Table~\ref{table:error_model}).

\subsection{Noise models}\label{sec:noise_models}

We consider two different types of noise models: a circuit-level depolarising noise model, and a circuit-level independent noise model. 
The depolarising noise model is widely used in the literature, and is useful for comparing to previous work.
Later we will consider biased noise, for which we use the independent noise model.

The circuit-level depolarising noise model is the same as that used in Refs.~\cite{chamberland2020topological, huang2020fault}, and is parameterised by a single variable $p$. Ancilla state preparation and measurement errors each occur with probability $2p/3$. With probability $p$, each CNOT gate is followed by a two-qubit Pauli error drawn uniformly from $\{I,X,Y,Z\}^{\otimes 2}\setminus I\otimes I$. A single qubit Pauli error drawn uniformly from $\{X,Y,Z\}$ occurs with probability $p$ after each idle single qubit gate location. Note that many of our syndrome extraction circuits are fully parallelised, and do not contain single qubit gates or idle locations.

In our circuit-level independent noise model, $Z$-type errors and $X$-type errors are independent. For a given error probability parameterised by $p_0$, we choose a high-rate error probability for $Z$-type errors $p_Z=p_0\eta/(\eta+1)$
and the low-rate error probability
$p_X=p_0/(\eta+1)$
for $X$-type errors. The bias $\eta=p_Z/p_X$ parameterises the relative strengths of $Z$-type and $X$-type errors. The total probability of any error is:
\begin{align}\label{eq:biased_total_error_rate}
\begin{split}
p_{tot}&=1-(1-p_X)(1-p_Z)\\
&=p_0-\frac{p_0^2\eta}{(\eta+1)^2}.
\end{split}
\end{align}
Each with probability $p_Z$, a CNOT gate is followed by an error in $\{IZ, ZI, ZZ\}$, chosen uniformly at random, an $X$-type ancilla is prepared or measured in an orthogonal state, and a single qubit idle for one time step undergoes a $Z$ error. Similarly, each with probability $p_X$, a CNOT gate is followed by an error randomly chosen from $\{IX,XI,XX\}$, a $Z$-type ancilla is prepared or measured in an orthogonal state, and a single qubit idle for one time step undergoes an $X$ error. Biased noise models are common in many physical realisations of quantum computers, and bias-preserving CNOT gates can be realised using stabilized cat qubits~\cite{puri2020bias}. We note that our techniques significantly improve performance even for small finite bias ($\eta\leq10$), which may be achievable even with CNOT gates that do not fully preserve bias, as is the case in many architectures~\cite{aliferis2008fault,guillaud2019repetition}.

\begin{table}
\renewcommand{\arraystretch}{1.2}
\begin{tabular}{ccccccccc}
Error type   & $G_1^X$  & $G_2^X$ & $G_1^Z$ & $G_2^Z$ & $P_X$ & $M_X$ & $P_Z$ & $M_Z$ \\
\hline
Depolarising & $\frac{4}{15}p$ & $\frac{8}{15}p$ & $\frac{4}{15}p$ & $\frac{8}{15}p$ & $\frac{2}{3}p$ & $\frac{2}{3}p$ & $\frac{2}{3}p$ & $\frac{2}{3}p$ \\
Independent & $\frac{1}{3}p_X$ & $\frac{2}{3}p_X$ & $\frac{1}{3}p_Z$ & $\frac{2}{3}p_Z$ & $p_X$ & $p_X$ & $p_Z$ & $p_Z$ \\
Ref.~\cite{bravyi2013subsystem} & $\frac{1}{4}p$ & $\frac{1}{2}p$ & $\frac{1}{4}p$ & $\frac{1}{2}p$ & $p$ & $p$ & $p$ & $p$ \\
\end{tabular}
\caption{The probability of a fault occurring for each type of circuit element under the two error models considered in this work, as well as for the depolarising error model used in Ref.~\cite{bravyi2013subsystem} for reference.}\label{table:error_model}
\end{table}

The probability of each different type of circuit element undergoing a fault for our two error models (as well as the error model in Ref.~\cite{bravyi2013subsystem} for comparison) is given in Table~\ref{table:error_model}.

\section{Numerical Analysis}\label{sec:numerical_analysis}

For all of the numerical results in this section, we used a local variant of the minimum-weight perfect matching (MWPM) decoder, described in \Cref{app:local_matching} and available at~\cite{higgott2021pymatching}, along with the Blossom V implementation of the Blossom algorithm~\cite{edmonds1965paths,kolmogorov2009blossom}.

\subsection{Subsystem toric codes}

\subsubsection{Gauge-fixing for depolarising noise}\label{sec:gauge_fixing_depol}

\begin{figure}
\includegraphics[width=0.8\columnwidth]{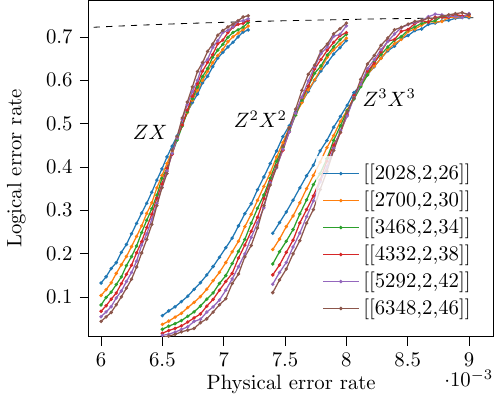}
\caption{Threshold plots for subsystem toric codes using a $(ZX)^{92}$ schedule (left), $(Z^2X^2)^{46}$ schedule (middle) and $(Z^3X^3)^{31}$ schedule (right) using a depolarising noise model.}
\label{fig:subsystem_toric_thresholds}
\end{figure}

We will now show how gauge-fixing can be used to improve the quantum error correcting performance of the subsystem toric code under a depolarising noise model. For this unbiased noise, we have used \textit{balanced} schedules, which we define to be of the form $Z^aX^a$ for some $a\in \mathbb{Z}^+$. We find that schedules that allow gauge-fixing increase the threshold from $0.666(1)\%$ for the standard $(ZX)^r$ schedule used in Ref.~\cite{bravyi2013subsystem} to $0.811(2)\%$ for the $(Z^4X^4)^r$ schedule, where gauge operators are fixed for three in every four rounds of measurements. In \Cref{fig:subsystem_toric_thresholds}, we show the thresholds for the $ZX$, $Z^2X^2$ and $Z^3X^3$ schedules. We see that both the $Z^2X^2$ and $Z^3X^3$ schedules are higher than the standard $ZX$ schedule, but the crossing is at a higher logical error rate. For these balanced schedules $(Z^aX^a)^r$ under depolarising noise, we find that $a=4$ is optimal (see Table~\ref{table:toric_depolarising_thresholds}). Therefore, schedule-induced gauge fixing makes the threshold of the subsystem toric code under depolarising noise much more competitive with the rotated surface code, which we find has a threshold of around 0.97\% under the same noise model and assumptions (state preparation and measurement each take half the time of a CNOT, and the logical error rate per time step is used). However, in \Cref{sec:biased_noise_numerics} we show that schedule-induced gauge fixing with the subsystem toric code can be used to outperform the rotated surface code for small finite bias $\eta>2.3$.

By using gauge fixing (setting $a>1$) we reduce the average stabiliser weight in the 3D matching graph, since the stabilisers introduced from gauge fixing have weight 3. The mean stabiliser weight in the 3D ($X$ check) matching graph for a $\{2c,4\}$ subsystem surface or hyperbolic code using a $(Z^qX^a)^r$ schedule (for any $q\geq 1$ or $r\geq 1$) is given by $3ca/(c(a-1)+1)$. So for the subsystem toric code ($c$=2), the mean stabiliser weights for the $(ZX)^r$, $(Z^2X^2)^r$ and $(Z^3X^3)^r$ schedules are 6, 4 and 3.6 respectively. We also reduce the average degree of vertices in the matching graph. For $a=1$ the mean vertex degree is 14, whereas for $a>1$, the mean vertex degree is $8ca/(c(a-1)+1)$, and so the $(ZX)^r$, $(Z^2X^2)^r$ and $(Z^3X^3)^r$ schedules have mean vertex degrees of 14, 32/3 and 9.6 respectively for the subsystem toric code. More properties of matching graphs for some homogeneous schedules with the subsystem toric code are given in Table~\ref{table:lattice_properties}.

\begin{table}
\begin{tabular}{ccccccc}
Schedule & $\bar{|s|}$ & $|s|_{\max}$ & $|s|_{\min}$ & $\bar{d}$ & $\Delta$ & $\delta$ \\
\hline
$Z^qX$ & 6 & 6 & 6 & 14 & 14 & 14 \\
$Z^qX^2$ & 4 & 6 & 3 & 10.67 & 16 & 8 \\
$Z^qX^3$ & 3.6 & 6 & 3 & 9.6 & 16 & 8 \\
$Z^qX^5$ & 3.33 & 6 & 3 & 8.89 & 16 & 8 \\
$Z^qX^{10}$ & 3.16 & 6 & 3 & 8.42 & 16 & 8 \\
\end{tabular}
\caption{The mean $\bar{|s|}$, maximum $|s|_{\max}$ and minimum $|s|_{\min}$ stabiliser weight and mean $\bar{d}$, maximum $\Delta$, and minimum $\delta$ degree of the $X$-check 3D matching graphs for various homogeneous schedules with the subsystem toric code.}\label{table:lattice_properties}
\end{table}

While we expect that reducing the average stabiliser weight and vertex degree in the matching graph should improve the threshold, increasing $a$ in balanced $Z^aX^a$ schedules also alters the edge fault probabilities. In time steps where gauge operators are fixed, $r_Z=0$ in Table~\ref{table:edge_faults}, reducing the edge weights for some edges of type 0, 1 and 2. However, in the time steps where gauge operators are not fixed, $r_Z=a$, and so increasing $a$ also increases the edge-fault probability for these edges of type 0, 1 and 2. Therefore, increasing $a$ increases the proportion of time steps where a space-like slice of the matching graph is a degree-3 hexagonal lattice with small edge fault probabilites, but also \textit{increases} the edge fault probabilities for the remaining time steps where the matching graph is \textit{not} fixed, and is instead a degree-6 triangular lattice. There is therefore a trade-off between increasing the edge weights, and decreasing the stabiliser weights and vertex degrees, and the $a=4$ schedule is the optimal compromise for schedules of the form $(Z^aX^a)^r$ for a circuit-level depolarising noise model.

\begin{figure}
\includegraphics[width=0.49\columnwidth]{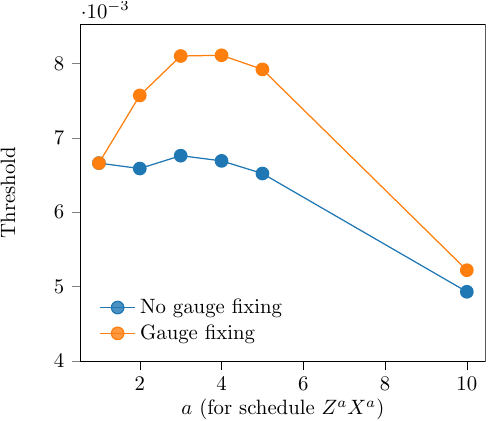}
\includegraphics[width=0.49\columnwidth]{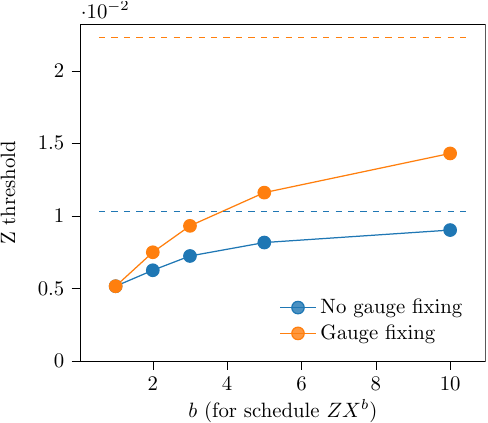}
\caption{Left: Circuit level depolarising threshold as a function $a$ for schedules of the form $Z^aX^a$, with and without gauge fixing. Right: $Z$ thresholds as a function of $b$ for schedules of the form $ZX^b$, both with (orange) and without (blue) gauge fixing, using a circuit-level independent noise model. The orange and blue dashed lines are the threshold achievable under infinite bias (using an $X$ schedule) with and without gauge fixing respectively. Error bars are smaller than the marker size and have been omitted for clarity.}
\label{fig:schedule_vs_threshold}
\end{figure}

Since changing the schedule alters both the matching graph via gauge fixing, as well as the edge fault probabilities, we can better understand how these two factors contribute to performance by studying them separately. In \Cref{fig:schedule_vs_threshold} we plot the threshold as a function of $a$ for balanced schedules $Z^aX^a$ both with and without using gauge fixing. The thresholds that do not use gauge fixing are decoded by always merging gauge factors of a stabiliser into a single vertex in the matching graph, even in time steps where they could be split (gauge factors fixed) using the techniques we have introduced. We see that for schedules that do not use gauge fixing, there is almost no improvement for $a>1$, with performance degrading for $a>4$. This demonstrates that almost all the improvement in threshold for depolarising noise is due to the use of gauge fixing, rather than the change in the noise model induced by the different schedule alone.

\subsubsection{Tailoring the 3D matching graph to biased noise using gauge fixing}\label{sec:biased_noise_numerics}

\begin{figure}
\includegraphics[width=0.8\columnwidth]{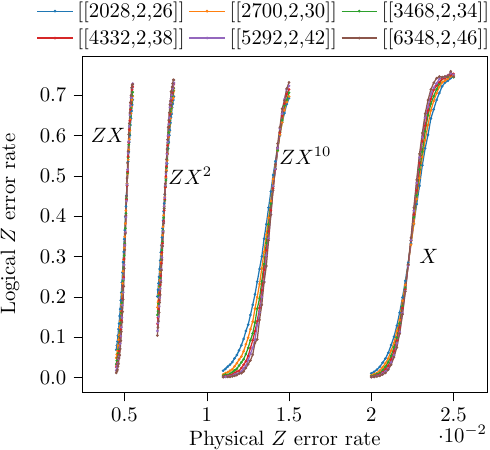}
\caption{$Z$ thresholds for unbalanced schedules of the form $ZX^{b}$, as well as an $X$ schedule, which gives an upper bound on the $Z$ threshold achievable using unbalanced schedules.}\label{fig:z_thresholds_unbalanced}
\end{figure}

By using \textit{unbalanced} schedules, where $X$ check operators are measured more frequently than $Z$ check operators (or vice versa), we can use gauge fixing to improve performance under biased noise models. Since we correct $X$ errors and $Z$ errors independently, we can define the $Z$ threshold $p_Z^{th}$ and $X$ threshold $p_X^{th}$ as the threshold for only $Z$-type or only $X$-type errors respectively.  In \Cref{fig:z_thresholds_unbalanced} we plot the $Z$ threshold for the unbalanced $ZX$, $ZX^2$, $ZX^{10}$ and $X$ schedules, under the independent circuit-level noise model. Increasing the ratio of $X$ checks to $Z$ checks significantly increases the $Z$ threshold from 0.52\% for the $ZX$ schedule up to 2.22\% for the $X$ schedule, which sets an upper bound.

By measuring $X$ checks more frequently, we also reduce the noise on data qubits caused by the CNOT gates used to measure $Z$ checks. To determine how much of the improvement in threshold comes from this reduced noise in the measurement schedule compared to the use of gauge fixing in the matching graph, we determine the thresholds both with and without using gauge fixing in \Cref{fig:schedule_vs_threshold}. We see that even without using gauge fixing, increasing the ratio of $X$ checks to $Z$ checks increases the $Z$ threshold, as expected. However, gauge fixing significantly boosts the $Z$ threshold further, and even a $ZX^5$ schedule using gauge fixing outperforms the best achievable $Z$ threshold without gauge fixing (using the $X$ schedule).

However, by increasing the ratio of $X$ to $Z$ checks, we also reduce the $X$ threshold of the code, which we must take into account when determining the total threshold under biased noise models. We now ask what the threshold is under the biased independent circuit-level noise model described in Section~\ref{sec:noise_models}, with bias parameter $\eta$. Specifically, for a given $\eta$, we wish to find the total physical error rate $p_{total}^{th}$ below which the total logical error probability $p_{total}^{log}$ of both logical $\bar{X}$ or $\bar{Z}$ errors vanishes as the distance $L$ of the code increases to infinity. A sufficient and necessary condition for a total error probability $p_{total}^\prime$ to be below the accuracy threshold for a decoder that decodes $Z$ and $X$ errors independently is that the probability of a $Z$-type error $p_{Z}^\prime$ be below $p_Z^{th}$ and the probability of an $X$-type error $p_{X}^\prime$ be below $p_X^{th}$.

The total error probability $p_{total}^{Z_{th}}$ when $p_Z=p_Z^{th}$ is
\begin{equation}\label{eq:tot_z_thresh}
p_{total}^{Z_{th}}=p_Z^{th} + p_Z^{th}(1-p_Z^{th})\frac{1}{\eta}
\end{equation}
and the total error probability $p_{total}^{X_{th}}$ when $p_X=p_X^{th}$ is
\begin{equation}\label{eq:tot_x_thresh}
p_{total}^{X_{th}}=p_X^{th}+p_X^{th}(1-p_X^{th})\eta.
\end{equation}
The total threshold $p_{total}^{th}$ is therefore given by
\begin{equation}\label{eq:tot_thresh}
p_{total}^{th}=\min(p_{total}^{Z_{th}},p_{total}^{X_{th}}).
\end{equation}

\begin{figure}[h]
    \centering
    \includegraphics[width=0.8\columnwidth]{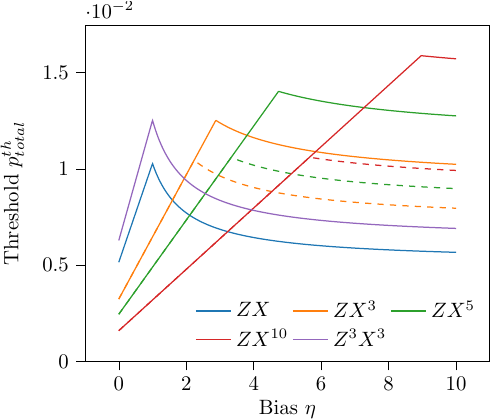}
    \caption{Threshold $p_{total}^{th}$ (see Eq.~\ref{eq:tot_thresh}) as a function of bias for different homogeneous schedules and under a circuit-level independent noise model. Dashed lines use the same schedule as the corresponding solid line of the same colour, except gauge fixing is not used, for the purpose of comparison.}
    \label{fig:schedule_thresholds_vs_bias}
\end{figure}

In \Cref{fig:schedule_thresholds_vs_bias} we plot $p_{total}^{th}$ as a function of the bias parameter $\eta$ for the subsystem toric code, and for a few different choices of homogeneous schedule. For the $ZX$ schedule, used in Ref.~\cite{bravyi2013subsystem}, and the $Z^3X^3$ schedule with gauge fixing, the optimal bias is $\eta=p_Z/p_X=1$. This is as expected, since the $X$ threshold is identical to the $Z$ threshold for these symmetric schedules. From Eqs.~\ref{eq:tot_z_thresh}, \ref{eq:tot_x_thresh} and \ref{eq:tot_thresh} we see that at $\eta=0$ and $\eta=\infty$ the total threshold is simply the $X$ threshold and $Z$ threshold, respectively. 

For each of the schedules for which $p_{total}^{th}$ is plotted in \Cref{fig:schedule_thresholds_vs_bias}, there are two regimes: to the left and to the right of the peak. To the left of the peak, the threshold is limited by the $X$ threshold, and is therefore given by \Cref{eq:tot_x_thresh}, which is linear in~$\eta$. 
To the right of the peak, the threshold is limited by the $Z$ threshold, and is therefore given by \Cref{eq:tot_z_thresh}, which is linear in~$1/\eta$. 
The optimal $\eta$ for a given schedule can be found by setting $p_{total}^{Z_{th}}=p_{total}^{X_{th}}$.

Even for small finite bias, using unbalanced schedules and gauge fixing significantly improves the total threshold compared to the traditional $ZX$ schedule, with a $2.8\times$ increase in threshold at $\eta=9$. With infinite bias the threshold rises to $2.22\%$ which is $4.3\times$ higher than the threshold of $0.52\%$ using standard $ZX$ schedule.
Each dashed line in \Cref{fig:schedule_thresholds_vs_bias} uses the same schedule as the corresponding solid line of the same colour, but without using gauge fixing to decode. 
For high bias, we see that approximately half of the improvement over the $ZX$ schedule can be attributed to the effect the new schedule has on the noise model, with the remainder attributed to the extra information used by gauge fixing when decoding.

For the rotated surface code, using the same schedule as in Ref.~\cite{brown2019handling}, we find a threshold under circuit-level independent noise of 0.741(2)\%. Therefore, the subsystem toric code (with a $ZX^3$ schedule and using gauge fixing) outperforms the rotated surface code for biases $\eta>2.3$.

Note that, for all the thresholds we have reported so far, we have used fully parallelised schedules. Whereas the $ZX$ schedule is fully parallelised with only $n_a=1$ ancilla qubits per triangle operator, the unbalanced $ZX^b$ schedules require two ancilla qubits per $X$ check operator ($n_a=1.5$), and the balanced $Z^aX^a$ schedules require two ancilla qubits per $X$ check operator \textit{and} per $Z$ check operator ($n_a=2$). Since there are $4n_a/3$ ancilla qubits per data qubit, this leads to a larger qubit overhead when using gauge fixing with parallelised schedules. We can choose not to parallelise the schedules, and instead simply omit gates in the $ZX$ schedule to construct our other schedules (e.g.~an unparallelised $ZX^2$ schedule can be constructed by omitting every other $Z$ measurement in the $ZX$ schedule). These schedules incur no qubit overhead, but instead introduce idle errors. The threshold with infinite bias using an unparallelised $X$ schedule is 1.25\%, compared to 2.22\% using a parallelised $X$ schedule, both an improvement over the 0.52\% threshold using the $ZX$ schedule. Near the threshold, using additional ancillas is clearly worthwhile, whereas far below threshold it may be beneficial to use an unparallelised schedule, using the additional qubits to instead construct a code with a larger distance.

To analyse the performance below threshold, we compare a $ZX$ schedule to an unparallelised $ZX^3$ schedule ($n_a=1$) using the $L=26$ subsystem toric code, both with and without using gauge fixing to decode. When using gauge fixing, the logical $Z$ error rate is reduced by around four orders of magnitude compared to the $ZX$ schedule. Without using gauge fixing, the logical error rate with the unparallelised $ZX^3$ schedule is slightly worse than with the $ZX$ schedule, since idle qubit errors are worse than qubit errors in the standard depolarising noise model~\cite{stephens2014fault}.

\begin{figure}
\includegraphics[width=0.8\columnwidth]{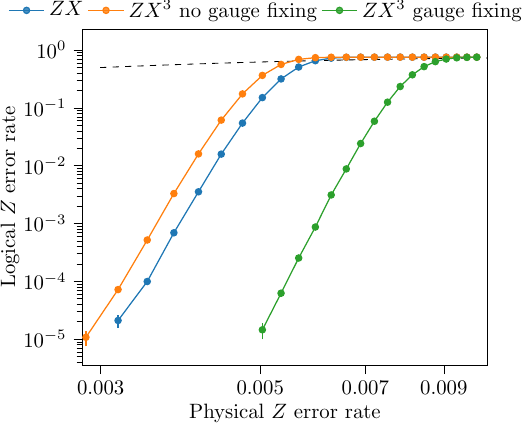}
\caption{Logical $\bar{Z}$ error rate of the $[[2028,2,26]]$ subsystem toric code using a $(ZX)^{36}$ schedule, as well as a $(ZX^3)^{12}$ schedule (using only a single ancilla by introducing idle time steps) with and without using gauge fixing in the matching graph. All schedules use 144 time steps, and the independent circuit-level noise model was used. The dashed black line is the probability that either of two physical qubits will suffer a $Z$ error during 144 time steps without using error correction.}
\end{figure}

\subsection{Performance of the finite-rate LDPC subsystem codes}\label{sec:subsystem_semi_hyperbolic_performance}

\begin{figure}
\includegraphics[width=0.8\columnwidth]{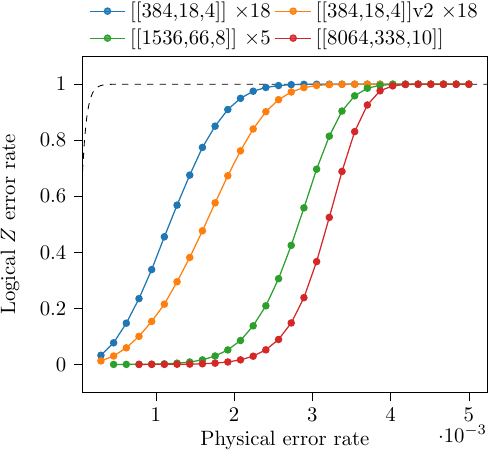}
    \caption{Performance of the extremal~$l=2$~$\{8,4\}$ subsystem semi-hyperbolic codes under a circuit-level depolarising noise model. Here, we fix the number of logical qubits to at least 338 for all codes, by using multiple copies of the smaller codes. A homogeneous~$(ZX)^{20}$ schedule is used for all codes, and the $y$ axis is the probability that at least one logical $Z$ error occurs. The dashed black line is the probability of a $Z$ error occurring on at least one of 338 physical qubits without error correction under the same error model for the same duration (80 time steps). For each code that encodes $k<338$ logical qubits, we use $m=\lfloor k/338\rfloor$ copies and plot the failure rate as~$p_{log}^*=1-(1-p_{log})^{m}$.}
    \label{fig:semi_hyperbolic_l2_fix_num_logical}
\end{figure}

We have simulated the performance of $l=2$ $\{8,4\}$ subsystem semi-hyperbolic codes under the circuit-level depolarising noise model. We are interested in finding the threshold value below which the logical error rate per logical qubit tends to zero as the code distance tends to infinity. Since the number of logical qubits $k$ increases with distance for this family of finite-rate codes, we fix the number of logical qubits by using multiple independent copies of the smaller codes. In \Cref{fig:semi_hyperbolic_l2_fix_num_logical} we plot the probability that at least one of 338 logical qubits suffers a $Z$ failure as a function of the depolarising error rate $p$. The [[8064,338,10]] code has the lowest logical error rate per logical qubit for physical error rates below 0.42\%, from which we conclude that the threshold is at least 0.42\%. We have not been able to obtain an upper bound on the threshold, since all codes have an error rate (per 338 logical qubits) of one for physical error rates above 0.42\%, within the precision provided by our numerical experiments.

\begin{figure}
\includegraphics[width=0.8\columnwidth]{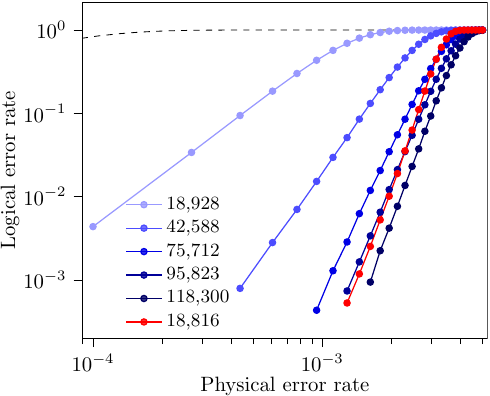}
\caption{Comparison of the [[8064,338,10]] l=2 $\{8,4\}$ subsystem semi-hyperbolic code (red), which has 8,064 data qubits and 10,752 ancillas, with $L=4,6,8,9$ and 10 subsystem toric codes (shades of blue), using a $(ZX)^{20}$ schedule (no gauge fixing) and a circuit-level depolarising noise model. We fix the number of logical qubits by plotting the probability that at least one of 169 independent copies of the subsystem toric codes suffers a logical $Z$ failure (i.e.~we plot $1-(1-p_{log})^{169}$ for the subsystem toric codes where $p_{log}$ is the probability that a single copy of the code suffers a logical $Z$ error). The total number of physical qubits (including ancillas) is given in the legend. The black dashed line is the probability that at least one of 338 physical qubits would suffer a $Z$ failure without error correction over the same duration.}\label{fig:compare_l2_semi_vs_toric}
\end{figure}

We now analyse the performance of the [[8064,338,10]] $l=2$ $\{8,4\}$ subsystem semi-hyperbolic code, which has the best ratio $n/(kd^2)=0.24$ of the codes we have constructed. In \Cref{fig:compare_l2_semi_vs_toric} we compare its performance with that of the $L=4,6,8,9$ and 10 subsystem toric codes. We use 169 independent copies of the subsystem toric codes, in order to keep the number of logical qubits (338) constant, and the total number of physical qubits used (including ancillas) is given in the legend. We find that the [[8064,338,10]] subsystem semi-hyperbolic code (which uses 18,816 physical qubits), outperforms the $L=4$ subsystem toric code (which uses 18,928 physical qubits to encode 338 logical qubits) by around three orders of magnitude at $p=0.15\%$. At a physical error rate of 0.2\% the performance of the [[8064,338,10]] subsystem semi-hyperbolic code is similar to the $L=9$ subsystem toric code, which uses 95,823 physical qubits to achieve the same logical error rate. This demonstrates that the [[8064,338,10]] subsystem semi-hyperbolic code requires $5.1\times$ fewer resources to achieve the same level of protection that the subsystem toric code would provide at a physical error rate of 0.2\%. 

\begin{figure}
\includegraphics[width=0.8\columnwidth]{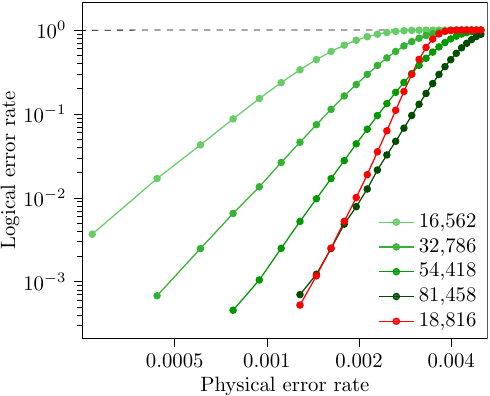}
\caption{Comparison of the [[8064,338,10]] l=2 $\{8,4\}$ subsystem semi-hyperbolic code (red), with $L=5,7,9$ and 11 rotated surface codes (shades of green), using a $(ZX)^{20}$ schedule (no gauge fixing) for the subsystem semi-hyperbolic code and a $(ZX)^{16}$ schedule for the rotated surface codes (both schedules require 80 time steps). We use a circuit-level depolarising noise model. We fix the number of logical qubits by plotting the probability that at least one of 338 independent copies of the rotated surface code suffers a logical $Z$ failure. The legend gives the total number of qubits (ancilla and data qubits) used. The black dashed line is the probability that at least one of 338 physical qubits would suffer a $Z$ failure without error correction over the same duration.}\label{fig:compare_l2_semi_vs_rotated}
\end{figure}

We also compare the [[8064,338,10]] subsystem semi-hyperbolic code with the rotated surface code, which is the leading candidate for realising fault-tolerant quantum computation, and has the optimal ratio $n/d^2=1$ for surface codes~\cite{bombin2007optimal}. This comparison is shown in \Cref{fig:compare_l2_semi_vs_rotated}, where we again keep the number of logical qubits fixed by using 338 independent copies of the rotated surface codes. At a circuit-level depolarising error rate of 0.15\%, the subsystem semi-hyperbolic code, using 18,816 physical qubits, has a similar performance to $L=11$ rotated surface codes using 81,458 physical qubits, a $4.3\times$ reduction in qubit overhead. We also compare the performance of the [[8064,338,10]] subsystem semi-hyperbolic code with a distance 6 rotated surface code, which has a slightly lower encoding rate (including ancillas), and find that the subsystem semi-hyperbolic code has a lower logical error rate below $0.43\%$.

To the best of our knowledge, the rotated surface code is the best performing code in the literature in terms of qubit overhead in the regime of around 0.15\% to 0.2\% circuit-level depolarising noise, which is roughly the same physical error rate assumed for practical implementations of fault-tolerant quantum computing~\cite{gidney2019factor,babbush2018encoding}. Since our subsystem semi-hyperbolic codes have a qubit overhead that is $4.3\times$ smaller than the rotated surface code at $p=0.15\%$, as demonstrated in \Cref{fig:compare_l2_semi_vs_rotated}, we therefore believe that they outperform all known quantum error correcting codes in terms of qubit overhead in this regime.

\begin{figure}
\includegraphics[width=0.8\columnwidth]{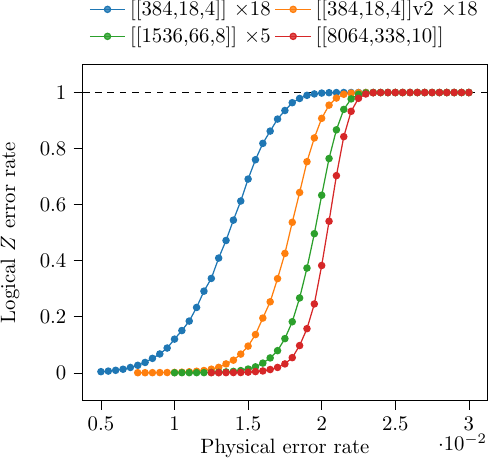}
\caption{Performance of the extremal subsystem $\{8,4\}$ $l=2$ semi-hyperbolic codes under a circuit-level independent noise model and using an $X$ schedule.}
\label{fig:semi_hyperbolic_l2_fix_num_logical_infinite_bias}
\end{figure}

Furthermore, we can use schedule-induced gauge fixing for the subsystem hyperbolic and semi-hyperbolic codes just as we did for the subsystem toric code. In \Cref{fig:semi_hyperbolic_l2_fix_num_logical_infinite_bias} we plot the threshold of the $l=2$, $\{8,4\}$ subsystem semi-hyperbolic codes under the independent circuit-level noise model using an $X$ schedule, and find a threshold of at least 2.4\%, exceeding that of the subsystem toric code (2.22\%). This threshold sets an upper bound on the thresholds that can be achieved using gauge fixing under biased noise models, and we expect that large gains can still be found even for small finite bias, as we found for the subsystem toric codes.

\section{Broader applications of our techniques}\label{sec:broader_applications}
\subsection{Inhomogeneous schedules}

We have so far only considered homogeneous schedules, however sometimes it may be advantageous to use schedules that are \textit{inhomogeneous}, where check operators in different faces of the lattice are given different schedules. 

As an example, consider two different unparallelised $ZX^4$ schedules, which we call $L_0$ and $L_1$, obtained by omitting three quarters of the $Z$ check operator measurements in the $ZX$ schedule, and such that $L_1$ is identical to $L_0$ other than a lag of $4$ check operator measurements. A section of 8 rounds of $X$ check operator measurements for these schedules looks like
\begin{tabular}{|c||c|c|c|c|c|c|c|c|c|c|c|c|c|c|c|c|}
\hline
$(ZX)^8$ & $Z$ & $X$ & $Z$ & $X$ & $Z$ & $X$ & $Z$ & $X$ & $Z$ & $X$ & $Z$ & $X$ & $Z$ & $X$ & $Z$ & $X$ \\
\hline
 $L_0$ & & $X$ & & $X$ & & $X$ & $Z$ & $X$ & & $X$ & & $X$ & & $X$ & $Z$ & $X$ \\
\hline
 $L_1$ & & $X$ & $Z$ & $X$ & & $X$ & & $X$ & & $X$ & $Z$ & $X$ & & $X$ & & $X$ \\
\hline
\end{tabular}
where each column corresponds to a measurement round of either $X$-type or $Z$-type check operators. We can assign either the $L_0$ or $L_1$ schedule to each face of the planar subsystem surface code independently, since each schedule is a subset of the $ZX$ schedule, for which we have a consistent measurement circuit for every face. Let $G_0^X$ be the set of $X$ triangle operators in faces assigned the $L_0$ schedule, and let $G_1^X$ be the set of $X$ triangle operators in faces assigned the $L_1$ schedule. Note that in each round of $X$ check operator measurements, either $G_0^X$, $G_1^X$ or $G_0^X\cup G_1^X$ may be fixed. 

\begin{figure}
\includegraphics[width=0.32\columnwidth]{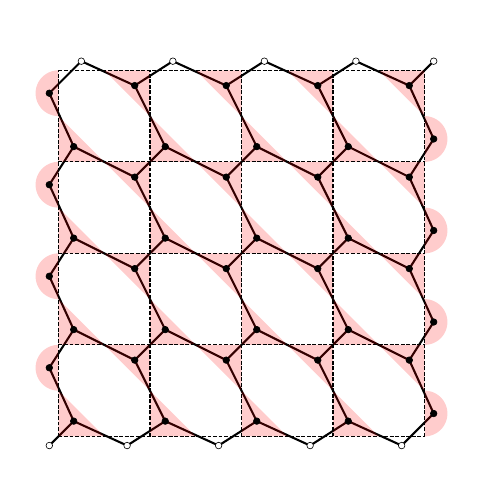}
\includegraphics[width=0.32\columnwidth]{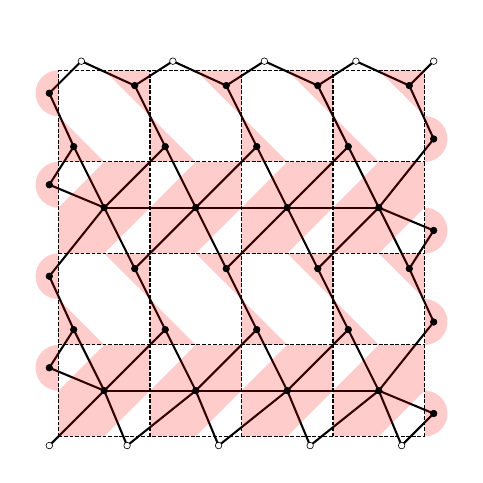}
\includegraphics[width=0.32\columnwidth]{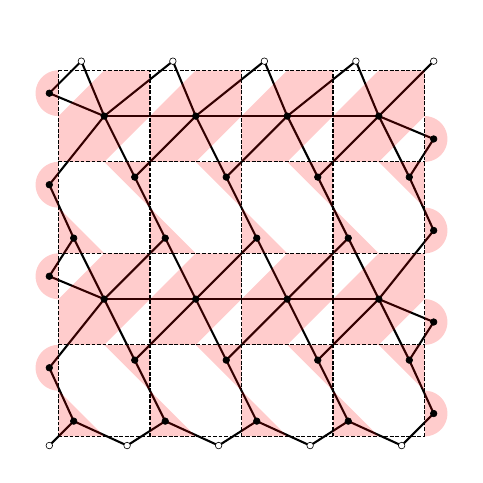}
\caption{Matching graphs ($X$-type) for the $L=5$ subsystem surface code with triangle operators fixed in all rows (left), odd rows (middle) and even rows (right). Filled and hollow circles correspond to stabilisers and boundary nodes respectively.}
\label{fig:inhomogeneous_matching_graphs}
\end{figure}

Can an inhomogeneous schedule be used to increase the $Z$ distance of a subsystem code? For the planar subsystem surface code, the only $Z$ logical is a Pauli $Z$ operator applied to each qubit in a column of the lattice, corresponding to a path in the matching graph joining the north and south boundaries. Consider the inhomogeneous schedule where we alternate between using the $L_0$ and $L_1$ schedule in each row of the lattice: we assign the schedule $L_{(i \mod 2)}$ to faces in the $i$\textsuperscript{th} row of the lattice. For a planar subsystem surface code with an odd distance, in each round of $X$ check operator measurements at least half of the gauge operators can be fixed: we can fix gauge operators in all rows, then in even rows, then all rows again, then odd rows, and so on in a cycle. In \Cref{fig:inhomogeneous_matching_graphs} we plot space-like slices (single time steps) of the 3D matching graph for when all rows, odd rows and even rows of gauge operators are fixed. Within each of these slices of the 3D matching graph, the shortest path between the north and south boundary is \textit{larger} than the $Z$ distance of the subsystem surface code itself. We expect that the shortest path between the north and south boundaries of the overall 3D lattice is also larger, leading to an increased $Z$ distance of $d_Z=\lfloor 3(L-1)/2\rfloor+1$, but do not prove this here. The $X$ distance cannot increase in this schedule, since none of the $Z$ gauge operators can be fixed.

Note that homogeneous schedules cannot increase the $Z$ or $X$ distance of the code, since there are always time steps where all $X$ gauge operators are measured simultaneously, as well as time steps where all $Z$ gauge operators are measured simultaneously. Measuring all $X$ gauge operators removes all $Z$ gauge operators from the stabiliser group, leaving time steps where none of the $Z$ gauge operators can be fixed (and therefore not increasing the $X$ distance), and similarly there are also time steps where no $X$ gauge operators can be fixed.

\subsection{Lattice surgery and code deformation}\label{sec:lattice_surgery}

\begin{figure}
\centering
\includegraphics{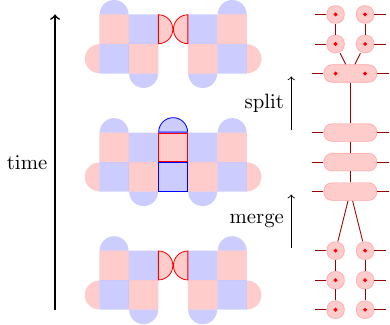}
\caption{A slice of the matching graph for lattice surgery, which can be interpreted as switching between different gauge fixes of a subsystem code. Left: the three stages of lattice surgery are shown for a distance 3 rotated surface code. Red (blue) squares and semi-circles denote $X$ ($Z$) stabilisers, with data qubits at their corners. Right: a slice of the matching graph for the $X$ stabilisers at the boundaries of the two codes where the merge takes place (denoted with red borders in the left diagram). Stabiliser measurements are repeated three times for each stage of lattice surgery, with the generalised difference syndrome used to connect the stabiliser with its gauge factors.}
\label{fig:lattice_surgery_matching_graph}
\end{figure}

It was shown in Ref.~\cite{vuillot2019code} that the techniques of lattice surgery~\cite{horsman2012surface} and code deformation~\cite{bombin2009quantum} can be interpreted as switching between different gauge fixes of a subsystem code. We can use this perspective to apply some of the techniques in this work to lattice surgery and code deformation. As an example, consider performing lattice surgery on two rotated surface code patches. During the merging step of lattice surgery, the weight two $X$ stabilisers on the opposing boundaries of the two patches are merged into weight 4 square stabilisers. These weight four stabilisers can be interpreted as stabilisers of a subsystem code, with the weight two checks that they are merged from being gauge operators of the subsystem code. This procedure is shown for distance 3 codes in the left side of \Cref{fig:lattice_surgery_matching_graph}, for which a single pair of weight two $X$ checks (with red borders) is merged into a single square stabiliser. Since each pair of these weight two boundary $X$ checks is a pair of gauge factors of the corresponding weight 4 stabiliser, we can use the merging and splitting technique given in Section~\ref{sec:gauge_fixing_matching} to construct the matching graph and decode them. This is shown on the right side of \Cref{fig:lattice_surgery_matching_graph}, where three repetitions are used for each of the three stages of lattice surgery. With this technique, each of the consecutive stages of lattice surgery can be connected using the generalised difference syndrome, leading to a single matching graph that can be used for error correction with the overlapping recovery method of Ref.~\cite{dennis2002topological}, and with information from the weight two boundary $X$ checks used directly where possible. The same ideas can also be readily applied to code deformation, which can also be viewed as gauge fixing of a subsystem code~\cite{vuillot2019code}, and involves merging surface code patches in a similar manner~\cite{bombin2009quantum}.

\subsection{Subspace codes from gauge fixing}

\begin{figure}
	\centering
	\includegraphics{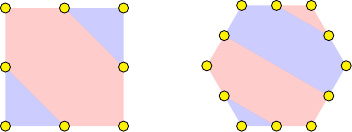}
	\caption{Gauge fixings of a square (left) and hexagonal (right) face of a subsystem toric and $\{6,4\}$ subsystem hyperbolic code, respectively. Yellow filled circles are data qubits, and $X$ and $Z$ stabilisers are denoted by red and blue filled polygons, respectively.}
	\label{fig:gauge_fix_face}
\end{figure}

Another use of gauge fixing is to derive families of subspace codes from subsystem surface, toric and hyperbolic codes, by choosing different abelian subgroups of the gauge group $\mathcal{G}$ to be the stabiliser group $\mathcal{S}$, permanently fixing some gauge operators as stabilisers. For example, by fixing all the $X$-type triangle operators in the subsystem toric code as stabilisers we obtain the hexagonal toric code, and by fixing $X$-type triangle operators in the $\{8,4\}$ subsystem hyperbolic code as stabilisers we obtain the $\{12,3\}$ hyperbolic code. 

By fixing different subsets of the triangle operators in the subsystem toric code, we can interpolate between the hexagonal toric code and its dual. To achieve this we define \textit{hexagonal surface density} codes, inspired by the surface-density and Shor-density codes of Ref.~\cite{li20192d}. To construct a (subspace) hexagonal surface density code with parameter $q_f$ from a subsystem toric code, we fix the $X$-type gauge operators in each face with probability $q_f$, else we fix the $Z$-type gauge operators. When $q_f=1$ we obtain the hexagonal surface code, and at $q_f=0$ we construct its dual, but setting $0<q_f<1$ allows us to interpolate between these two extremes. With $q_f=0.5$, there are both weight 6 and weight 3 $X$-type and $Z$-type stabilisers, and both $X$-type and $Z$-type stabilisers have average weight 4. The same idea can be directly applied to subsystem hyperbolic codes: applied to the $\{8,4\}$ subsystem hyperbolic code, we can interpolate between the $\{12,3\}$ hyperbolic code and its dual, for example. 

For the subsystem hyperbolic codes, we can choose to fix only a subset of the triangle operators within each face. Consider the code obtained by fixing a single $Z$ triangle operator (chosen at random) within each face of the $\{6,4\}$ subsystem hyperbolic code, as well as the single $X$ triangle operator that commutes with it (an example of this for a single face is shown in \Cref{fig:gauge_fix_face}). For both the $X$ and $Z$ stabilisers, half have weight 6, and the other half have weight 3. This hyperbolic code, derived from an \textit{irregular} lattice, has average stabiliser weight 4.5 for both $X$ and $Z$ stabilisers, an improvement on the weight 5 stabilisers in the $\{5,5\}$ hyperbolic code, which has the smallest stabiliser weight of hyperbolic codes derived from self-dual regular lattices.

We can also use our choice of abelian subgroup of the gauge group to tailor codes to spatially inhomogeneous noise models, where the noise is biased towards $Z$-type errors in some regions of the lattice, and biased towards $X$-type errors in other regions. We can fix $X$-type gauge operators in regions where there is a $Z$ bias, locally reducing the vertex degree and stabiliser weight in the $X$-type matching graph, and likewise we can fix $Z$-type gauge operators where there is $X$ bias. This method of tailoring a code to spatially inhomogeneous noise models has been demonstrated in Ref.~\cite{li20192d} using gauge fixes of the Bacon-Shor code, and the same ideas can be readily applied here to gauge fixes of subsystem surface, toric and hyperbolic codes.

\section{Conclusion}\label{sec:conclusion}
	
	In this work, we have introduced new techniques and constructions for quantum error correction that improve upon the widely-studied surface code in several ways.
	While the surface code requires four-qubit measurements and encodes a single logical qubit, we introduce families of quantum error correcting codes that use only three-qubit measurements and encode a number of logical qubits $k$ proportional to the number of physical qubits $n$.
	Furthermore, we have introduced a technique, which we call schedule-induced gauge fixing, that improves the performance of a wide class of codes, especially under biased noise models.
		
	Schedule-induced gauge fixing changes the order in which check operators are measured in subsystem codes.
	While the check operators of subsystem codes do not all mutually commute, we find that grouping blocks of mutually commuting check operators together allows us to obtain more useful information without increasing the total number of measurements.
	By making consecutive measurements of the same gauge operators they can be treated temporarily as stabilisers, and we introduce a method for decoding, based on minimum-weight perfect matching (MWPM), that takes advantage of this additional information.
	When applied to the subsystem surface code with three-qubit check operators, we can switch repeatedly between the hexagonal surface code and its dual, both of which are abelian subgroups of the gauge group of the code. 
	We find that the threshold under circuit-level depolarising noise can be increased from 0.67\% to 0.81\% by making four consecutive measurements of each gauge operator in the measurement schedule.
	The improvement is even more significant under biased noise models. With an independent $Z$-biased circuit-level noise model, $X$ check operators can be repeated (and fixed) more frequently, leading to an even higher threshold under small finite bias, up to 2.22\% under infinite bias.
	Below threshold, gauge fixing reduces the logical error rate by several orders of magnitude for biased noise models.

	Schedule-induced gauge fixing can be applied \textit{in software}, with no changes to the underlying hardware interactions necessary. 
	This allows both the code and the decoder to be tailored to the noise model even if it cannot be fully characterised prior to device fabrication.
	Furthermore, the decoding method only changes the structure of the matching graph, with no additional overhead in runtime, and other decoders such as Union-Find~\cite{delfosse2017almost,huang2020fault}, which has almost-linear runtime, can be directly substituted for MWPM in our procedure.
	
	The same techniques can also be directly applied to a broad class of subsystem codes beyond the subsystem surface code, including the Bacon-Shor code~\cite{bacon2006operator}, the heavy hexagon code~\cite{chamberland2020topological}, and some compass codes~\cite{li20192d}, and future work could investigate the performance improvements achievable using schedule-induced gauge fixing with these codes. 
	It would also be interesting to generalise the decoding method to other subsystem codes where syndrome defects do not come in pairs, such as the gauge colour code~\cite{bombin2015gauge}, amongst others~\cite{bombin2010topological,suchara2011constructions}.

	A drawback of subsystem codes is that they typically have a smaller encoding rate $k/n$ compared to their subspace counterparts. 
	To address this issue, we generalise the subsystem surface code to surfaces with negative curvature, constructing families of quantum LDPC subsystem codes with a finite encoding rate and only three-qubit check operators. 
	We call these codes subsystem hyperbolic and subsystem semi-hyperbolic codes, and show how the symmetry group of the tessellation can be used to construct check operator measurement circuits which require only four time steps to implement.
	Thanks to the weight-three check operators, these measurement circuits allow us to correct up to the full code distance fault-tolerantly.

	By simulating the performance of subsystem semi-hyperbolic codes under circuit-level depolarising noise, we find that they can require $4.3\times$ fewer physical qubits than the rotated surface code and $5.1\times$ fewer physical qubits than the subsystem toric code to achieve the same physical error rate at around $0.15\%$ to $0.2\%$. To the best of our knowledge, they therefore outperform all known quantum error correcting codes in terms of qubit overhead in this practical regime of circuit-level depolarising noise. Furthermore, these subsystem semi-hyperbolic codes belong to a family of codes that achieve distance scaling as $\sqrt{n}$, and that we expect to maintain a reduced qubit overhead relative to the surface code even at higher distances. 
	These codes are \textit{locally} Euclidean, which is encouraging for the prospect of physical implementations in modular architectures~\cite{nickerson2014freely,kalb2017entanglement,kollar2019hyperbolic}.
	
	We have also found a threshold of $0.42\%$ for the subsystem semi-hyperbolic codes under a circuit-level depolarising noise. 
	All of the techniques for schedule induced gauge-fixing that applied to the subsystem toric code can also be applied to subsystem semi-hyperbolic codes, and we find a threshold of $2.4\%$ under infinite bias, exceeding that of the subsystem toric code.
	
	Our work has focussed on reducing the qubit overhead of quantum error correction, however reducing the time overhead of implementing logical gates is also an important problem.
	In Ref.~\cite{breuckmann2017hyperbolic} it was shown how lattice surgery and Dehn twists can be used to implement logical gates in hyperbolic codes.
	While these techniques should generalise straightforwardly to the subsystem hyperbolic codes we have introduced, in the future it would be interesting to compare the time overhead of these methods with those used for surface codes, as well as to investigate alternative methods for implementing fault-tolerant logical operations.
		
	A key advantage of all the subsystem codes we have constructed and used in this work is that they all use check operators of weight three, compared to the weight-four stabilisers of the surface code. 
	Besides enabling bare-ancilla fault-tolerance and efficient measurement schedules, weight-three gauge operators can be helpful for handling leakage errors~\cite{brown2019handling}, and direct three-qubit parity check measurements have been proposed in Ref.~\cite{divincenzo2013multi}.  
	Since the average degree of the interaction graph is lower than the surface code, we also expect these codes to suffer from fewer frequency collisions and less crosstalk than the surface code in superconducting qubit architectures~\cite{chamberland2020topological}.
	On the other hand, if high-weight stabiliser measurements are available in hardware, then it may be possible to reduce the qubit overhead of our subsystem codes even further (likely at the cost of a lower threshold) by using a single ancilla qubit per stabiliser rather than per gauge operator, and measuring along the gauge operators to retain bare-ancilla fault-tolerance~\cite{li2018direct}.

	While there has been significant progress in the development of quantum LDPC codes with improved parameters $[[n,k,d]]$ relative to the surface code, our work provides the first evidence that these improvements can be retained even under circuit-level depolarising noise.
	We have demonstrated the advantages that can be had from co-designing an error correcting code along with its parity check measurement schedule.
	In particular, we have shown that subsystem codes offer great promise in reducing the weight of check operators in quantum LDPC codes, as well as enabling improved performance under biased noise models through the use of schedule-induced gauge fixing.
	Furthermore, our results show that symmetries present in the construction of quantum LDPC codes can also be crucial for optimising parity-check measurement schedules.
	We hope that our work will inspire the construction of new families of quantum LDPC codes designed using similar principles, further reducing the overhead of fault-tolerant quantum computation.

\begin{acknowledgments} 
	NPB would like to thank Steve Flammia for pointing out reference~\cite{bravyi2013subsystem}. 
	The authors would like to thank Andrew Landahl, Lingling Lao and Michael Newman for helpful discussions.
	We would also like to thank Dan Browne for feedback on our manuscript.
	OH acknowledges support from the Engineering and Physical Sciences Research Council [grant number EP/L015242/1].
	NPB is supported by his UCLQ Fellowship.
	The authors also acknowledge the use of the UCL Myriad High Performance Computing Facility (Myriad@UCL), and associated support services, in the completion of this work.
\end{acknowledgments}
	
\bibliography{bibliography}

\appendix

\section{Local matching decoder}\label{app:local_matching}

In order to reduce the complexity of the minimum-weight perfect matching (MWPM) decoder, we use an approximate version which we call \textit{local matching}. In our local matching decoder, we only check if each defect can be matched to one of the $k$ closest defects in the matching graph (rather than considering all other defects in the matching graph).

Given a matching graph $G$ (containing a vertex for each stabiliser measurement or boundary and a weighted edge for each single error) and a syndrome vector $\bf{z}$ (where $\mathbf{z}[i]=1$ if stabiliser $i$ is measured to be $-1$, otherwise $\mathbf{z}[i]=0$), the first step of a standard implementation of MWPM is to construct the \textit{defect graph} $V$, which contains a vertex for each defect $v$ (where by definition $\mathbf{z}[v]=1$) in $G$ and an edge for each possible pair of defects, weighted with the distance between them in $G$. Edmond's Blossom algorithm is then used to find a minimum-weight perfect matching in $V$~\cite{edmonds1965paths}, and the product of Pauli operators, each corresponding to a shortest path between matched pairs of defects, is returned as a correction.

In our local matching algorithm we include fewer edges in the defect graph $V$ than used in standard MWPM. For each defect $v$ in $G$, rather than finding the distance to every other defect using Dijkstra's algorithm, we instead find the distance to the $m$ nearest defects using our \textit{local Dijkstra} algorithm, and include $m$ edges to these defects in $V$, each weighted by their distance from $v$ in $G$. Pseudocode for the local Dijkstra algorithm, for finding the $m$ nearest neighbours of a defect $s\in G$, is given in Algorithm~\ref{alg:local_dijkstra}, which outputs a list $l$ of the $m$ nearest defects, their distances $d$ to $s$, and a predecessor list $p$. We use Kolmogorov's Blossom V implementation of MWPM in C++ to find the MWPM in $V$~\cite{kolmogorov2009blossom}.

\begin{algorithm}[h]
\Fn{LocalDijkstra($G$, $\mathbf{z}$, $m$, $s$)}{
For each $u\in G$, $d[u]=\infty$, $p[u]=u$\;
$d[s]=0$\;
Initialise priority queue $Q$\;
$Q.insert(s)$\;
Initialise empty list of found defects $l$\;
 \While{$Q$ is not empty and $length(l)<m$}{
  $u=Q.ExtractMin()$\;
  \If{$\mathbf{z}[u]=1$}{
  	$l.Insert(u)$\;
  }
  \For{each vertex $v$ adjacent to $u$ in $G$}{
  	\If{$weight(u,v)+d[u]<d[v]$}{
		$d[v]=weight(u,v)+d[u]$\;
		$p[v]=u$\;
		\eIf{$d[v]$ previously equal to $\infty$}{
			$Q.Insert(v)$\;
		}{
			$Q.DecreaseKey(v)$\;	
		}
	}
  }

  }
 }
 \caption{Local Dijkstra Algorithm}
 \label{alg:local_dijkstra}
\end{algorithm}

\begin{figure}
\includegraphics[width=0.7\columnwidth]{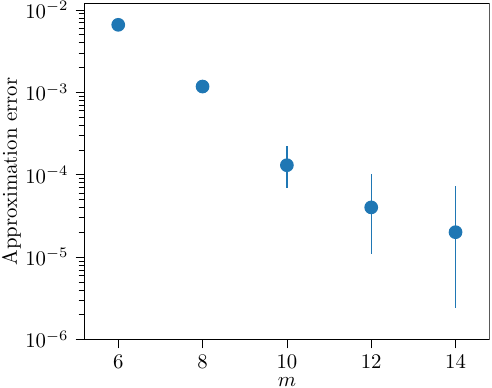}
\caption{The approximation error of the local matching decoder, defined as the fraction of trials for which the weight of the minimum weight perfect matching found by our local matching decoder differs from an exact minimum weight perfect matching. We use $10^5$ trials for each $m$, using an independent noise model with $p=6\%$ (and noiseless syndromes), for an $L=30$ toric code. Error bars are 95\% Clopper-Pearson binomial proportion confidence intervals. For $m=16$, $18$ and $20$ we also ran $10^5$ trials and found that the local matching was equivalent to exact matching for all trials.}
\label{fig:local_matching_approximation_error}
\end{figure}

Note that our local matching algorithm is similar to the strategy used in Ref.~\cite{fowler2012towards}, where defects are initially only matched with other defects that are within a given radius $r$ (determined by their coordinates in the 3D surface code matching graph). While this strategy is effective for codes derived from tessellations of Euclidean surfaces, it does not generalise well to the codes we have derived from tessellations of hyperbolic surfaces, where using coordinates as a proxy for finding the closest defects is less straightforward. Our approach only uses the structure of the matching graph itself, not the coordinates of defects, and therefore readily generalises to the codes we have derived from hyperbolic tessellations. Our adaptation of MWPM is also similar to the decoder used in Ref.~\cite{xu2019high}, where rather than restricting the number of neighbours of each vertex in $V$ as we do here, they instead impose a threshold on the maximum distance between vertices in $V$.

We have analysed the accuracy of our local matching decoder at approximating exact MWPM. In \Cref{fig:local_matching_approximation_error} we show that, empirically, the approximation error decreases exponentially with $m$. For $m>14$ we did not observe any differences in the weight of the matchings found by local matching and exact MWPM in any of the $10^5$ trials run on the $L=30$ toric code. Note that where our local matching differs from exact MWPM the solution given is still very good (low weight), so differences in the logical error rate between local matching and exact MWPM are likely far more rare than the differences in the exact minimum-weight matching solution measured here. We have used $m=20$ for all simulations in this work. Our implementation of the local matching decoder is available as a Python package at~\cite{higgott2021pymatching}.
	
\section{Tessellations of closed surfaces}\label{sec:tessellations}

\begin{figure}
	\centering
	\subfloat[]{
 	\includegraphics[width=0.47\columnwidth]{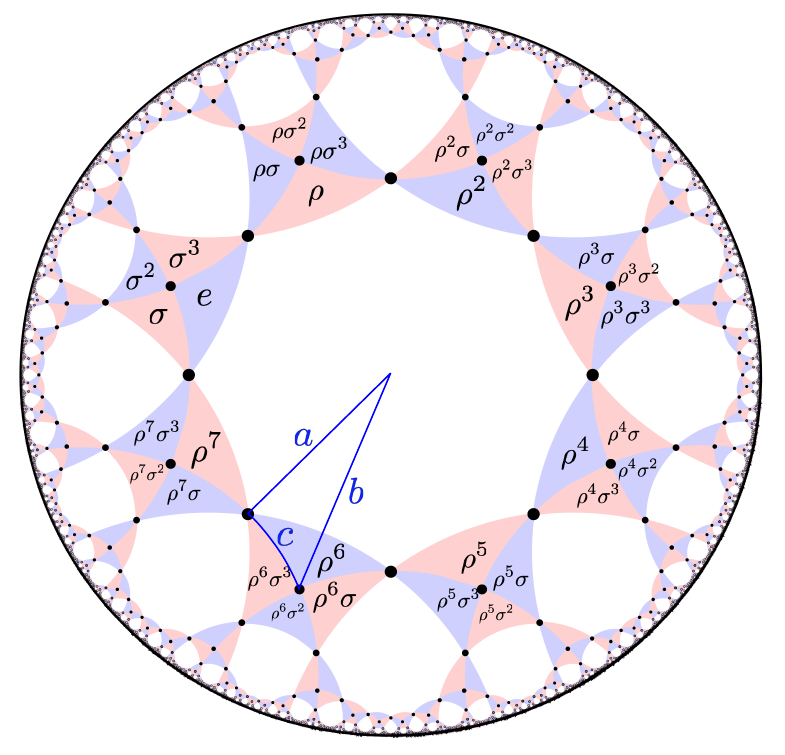}
	}
	~
	\subfloat[]{
	\includegraphics[width=0.47\columnwidth]{labelled_84_tessellation_no_isometries.pdf}
	}
	\caption{The $\{8,4\}$ subsystem hyperbolic code. (a) Each triangle operator can be uniquely identified with an element of the proper symmetry group $G_{8,4}^{H+}$ of the lattice (after identifying a triangle operator with the identity element $e$). We have labelled a fundamental triangle in blue. (b) Each triangle operator can be labelled with an element of $\mathbb{Z}_4$ using the homomorphism $h(\rho)=h(\sigma)=1$ from $G_{8,4}^{H+}$ to $\mathbb{Z}_4$. The neighbourhood of each triangle operator (the labels and relative locations of triangle operators it overlaps with) is the same as in the toric code.}\label{fig:subsystem_hyperbolic_homomorphism}
	\end{figure}

	We will now give some additional background on tessellations of closed Euclidean and hyperbolic surfaces, since these are used to construct the subsystem hyperbolic and semi-hyperbolic codes in this work. An $\{r,s\}$ tessellation subdivides a surface into disjoint faces, where each face is an $r$-gon, and $s$ faces meet at each vertex. Using \textit{Wythoff's kaleidoscopic construction}, an $\{r,s\}$-tessellation can be related to a symmetry group $G_{r,s}$ of distance-preserving maps (isometries). $G_{r,s}$ is generated by reflections on the edges of one of the $2r$ right triangles induced by the symmetry axes of a face ($r$-gon) of the tessellation. Each triangle has internal angles $\pi/2$, $\pi/r$ and $\pi/s$, and will from now on be referred to as a \textit{fundamental triangle}. In \Cref{fig:subsystem_toric_homomorphism}(a) and \Cref{fig:subsystem_hyperbolic_homomorphism}(a) we draw a fundamental triangle of the $\{4,4\}$ and $\{8,4\}$ tessellations respectfully, with sides labelled by the reflections $a$, $b$ and $c$ which act on them, and which generate $G_{r,s}$. Note that the isometries $a^2$, $b^2$, $c^2$, $(ac)^2$, $(ab)^r$ and $(ca)^s$ are equivalent to doing nothing and, since these are the only relations satisfied by $G_{r,s}$, the group has presentation
\begin{equation}
    G_{r,s}=\langle a,b,c |a^2=b^2=c^2=(ac)^2=(ab)^r=(bc)^s=e \rangle
\end{equation}
where $e$ is the identity element. By fixing one fundamental triangle as a fundamental domain of $G_{r,s}$, every other fundamental triangle can be labelled uniquely by an element of $G_{r,s}$. 

We will be constructing codes derived from $\{r,s\}$-tessellations of \textit{closed} Euclidean and hyperbolic surfaces. 
The process of defining a closed surface is called \textit{compactification}.
A regular tessellation of a closed surface can be defined by a quotient group $G^H_{r,s}\coloneqq G_{r,s}/H$, where $H$ is a finite index, normal subgroup of $G_{r,s}$ with no fixed points (see \cite{breuckmann2016constructions} for more details). 
Note that the generators of $H$ become relations in the presentation of $G_{r,s}/H$, so compactification can be interpreted as adding additional relations into the presentation of the symmetry group of the tessellation of the hyperbolic plane.
An important subgroup of $G_{r,s}$ is the proper symmetry group $G_{r,s}^+$ generated by double reflections, or \textit{rotations}, $\rho=ab$ and $\sigma=bc$. This group has presentation
\begin{equation}
G_{r,s}^+ = \langle \rho,\sigma | (\rho\sigma)^2=\rho^r=\sigma^s=e \rangle
\end{equation}
where $e$ is again the identity element. Regular tessellations of \textit{orientable} closed surfaces can be constructed from a quotient group $G_{r,s}^{H+}\coloneqq G_{r,s}^+/ H$, where $H$ is a normal subgroup of $G_{r,s}^+$.

	\section{Symmetry groups that admit subsystem hyperbolic codes}\label{app:subsystem_colouring}
	
	In \Cref*{sec:subsystem_hyperbolic_codes} of the main text we introduced subsystem hyperbolic codes, which are derived from $\{2c, 4\}$ tessellations of hyperbolic surfaces, where $c\in\mathbb{Z}^+$ and $c> 2$.
	In this section we will show how a subsystem hyperbolic code can be described in terms of the symmetry group of the tessellation from which it is derived.
	By doing so we will show what conditions must be satisfied by the compactification procedure for a $\{2c, 4\}$ tessellation of a closed hyperbolic surface to be used for constructing a subsystem hyperbolic code.

Let us first consider some properties of the subsystem toric code in group theoretic terms. 
These properties will later be used as requirements for the subsystem hyperbolic codes we define. 
First, note that each triangle operator (gauge generator) of the subsystem toric code can be identified by a \textit{pair} of fundamental triangles related by a $b$ reflection in $G_{4,4}$. 
In other words, each triangle operator is identified by a \textit{left coset} of the subgroup $\langle b\rangle$ given by $g\langle b\rangle\coloneqq \{g,gb\}$ for some $g\in G^H_{4,4}$, and thus each element $g\in G^H_{4,4}$ identifies a unique triangle operator (but not vice versa). 
For now we will consider only the Pauli type of each triangle operator, which can be either $Z$-type (blue) or $X$-type (red).
We will call an assignment of a Pauli type to each triangle operator a \textit{colouring}. 
For the subsystem toric code, note that blue triangle operators are always mapped to red triangle operators by either an $a$ or $c$ reflection, and vice versa. 
We will make this property a requirement of our subsystem hyperbolic codes, and will call a colouring that satisfies this property a \textit{valid} colouring. 

Since each triangle operator can be identified by the coset $g\langle b\rangle$ of an element $g\in G^H_{r,s}$, and after identifying each \textit{colour} of triangle operator with a different element of the cyclic group $\mathbb{Z}_2=\mathbb{Z}/2\mathbb{Z}$, a colouring of the triangle operators can be achieved by defining an appropriate function $f:G^H_{r,s}\rightarrow \mathbb{Z}_2$. 
The constraint that either $a$ or $c$ reflections map a triangle operator to another of a different type, with $b$ reflections leaving it invariant, defines the image of the generators and identity element $e$ of $G^H_{r,s}$ by~$f$ to be 
\begin{align}\label{eq:f_homomorphism}
\begin{split}
f(a)=f(c)&=1,\\
f(b)=f(e)&=0.
\end{split}
\end{align}

Since we require that, by definition of the code, the action of a reflection $a$, $b$ or $c$ should have the same effect on the colour of a triangle operator no matter which triangle operator we apply it to, this implies that $f(g_ig_j)=f(g_i)+f(g_j)\quad\forall g_i,g_j\in G^H_{r,s}$. This implies (from the definition of a homomorphism) that $f$ must extend to a \textit{group homomorphism} from $G^H_{r,s}$ to~$\mathbb{Z}_2$. For each triangle operator to be assigned a unique colour, we must also have that $f(r_i)=0$ for each relation $r_i$ in the presentation of $G^H_{r,s}$. 
This latter condition is in fact also a necessary and sufficient condition for the function $f$ to extend to a homomorphism from $G^H_{r,s}$ to~$\mathbb{Z}_2$~\cite{1402612}.
This constraint $f(r_i)=0$ holds not just for the $\{4,4\}$ tiling, but also $\{r,s\}$ tilings for which $r$ and $s$ are even, since $(ab)^r=e$ and $(bc)^s=e$ are relations. The constraints do not hold if either $r$ or $s$ are odd. However, we also have the constraint $f(g_i)=0$ on the generators $g_i$ of the normal subgroup $H$ defining the compactification (since these generators are relations in $G^H_{r,s}$) and, therefore, only a subset of the possible compactifications of these regular tessellations admit valid colourings.

We must also ensure that each triangle operator in a coloured tessellation commutes with every stabiliser, and that all stabilisers mutually commute (since by definition $\mathcal{S}$ is abelian and the center of $\mathcal{G}$). We will now show that this further restricts us to tessellations where $s=4$ faces meet at each vertex. For regular tessellations of closed Euclidean or hyperbolic surfaces, we are already restricted to $s\geq 3$, and we already require that $s$ be even to ensure a valid colouring. For all $s\in\{6,8,10,\ldots\}$ we see that each triangle operator anti-commutes with the stabiliser (of the opposite Pauli-type) belonging to the face related to it by a $(bc)^3$ rotation, since it overlaps with this stabiliser on only a single qubit. On the other hand, for $s=4$, it can be directly verified that each triangle operator commutes with all stabilisers, since each triangle operator overlaps on either zero or two qubits with stabilisers of the opposite Pauli type. Since stabilisers are products of non-overlapping triangle operators, all stabilisers must also mutually commute. We are therefore restricted to tessellations with $s=4$ faces meeting at each face and with $r=2c$ sides to each face, and for which $f(g_i)=0$ for each generator $g_i$ of the normal subgroup $H$ defining the compactification.

\section{Group theoretic condition for consistent scheduling}\label{app:labelling_homomorphism}

\begin{figure}
	\centering
	\subfloat[]{
 	\includegraphics[width=0.44\columnwidth]{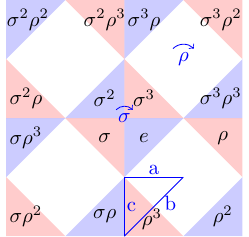}
	}
	~
	\subfloat[]{
	\includegraphics[width=0.42\columnwidth]{labelled_subsystem_toric.pdf}
	}
	\caption{An $L=2$ subsystem surface code. (a) After associating a triangle operator with the identity element $e$, every triangle operator is in one-to-one correspondence with an element of the proper symmetry group $G_{4,4}^{H+}$ of the tessellation. In blue we have labelled a fundamental triangle with sides $a$, $b$ and $c$, as well as the rotations $\rho=ab$ and $\sigma=bc$. (b) Each triangle operator can be labelled with an element of the cyclic group $\mathbb{Z}_4$ using the homomorphism $h(\rho)=h(\sigma)=1$ from $G_{4,4}^{H+}$ to $\mathbb{Z}_4$.}\label{fig:subsystem_toric_homomorphism}
	\end{figure}

In \Cref*{sec:scheduling_condition} of the main text, we showed that any translationally invariant schedule for the subsystem toric code assigns the same schedule to each triangle operator with the same \textit{label}, where a label is an assignment of an element of the cyclic group $\mathbb{Z}_4$ to each triangle operator as shown in \Cref{fig:subsystem_toric_homomorphism}(b). We will now describe this labelling of the triangle operators of the subsystem toric code in terms of the proper symmetry group $G_{r,s}^{H+}$ of orientation-preserving symmetries of the lattice, generated by the rotations $\rho$ and $\sigma$ (shown in \Cref{fig:subsystem_toric_homomorphism}(a)).
First note that, after choosing any triangle operator to be the fundamental domain, each triangle operator is now identified by a unique element in $G_{r,s}^{H+}$, and we will denote by $T_g$ the triangle operator identified by $g\in G_{r,s}^{H+}$.
A labelling of the triangle operators is then defined by a function $h:G_{r,s}^{H+} \rightarrow  \mathbb{Z}_4$.
Note that, for the labelling of the subsystem toric code in \Cref{fig:subsystem_toric_homomorphism}(b), applying either a $\rho$ or $\sigma$ rotation to \textit{any} triangle operator adds one (modulo 4) to the label.
Using similar arguments to those given in \Cref{app:subsystem_colouring} for valid colourings, we see that the function $h$ must extend to a homomorphism $h:G_{r,s}^{H+} \rightarrow  \mathbb{Z}_4$ with
\begin{equation}\label{eq:labelling_homomorphism}
h(\rho)=h(\sigma)=1.
\end{equation}

We can generalise a translationally symmetric schedule of the subsystem toric code to subsystem hyperbolic codes by first labelling the triangle operators of a subsystem hyperbolic code in such a way that the neighbourhood of each triangle operator is the same as it would be in the subsystem toric code, and then apply the same schedule to all triangle operators with the same label in the subsystem hyperbolic code. The \textit{neighbourhood} of a triangle operator $T$ is the relative position and label of the triangle operators that overlap with $T$ on at least one qubit (each of which we call a \textit{neighbour}). We see from \Cref{fig:subsystem_toric_homomorphism}(b) that each triangle operator $T_g$ in the subsystem toric code has seven neighbours: $T_{g\sigma}$, $T_{g\sigma^2}$, $T_{g\sigma^{3}}$, $T_{g\rho}$, $T_{g\rho\sigma}$, $T_{g\rho^{-1}}$ and $T_{g\rho^{-1}\sigma^{-1}}$. In the toric code, exactly three of these neighbours overlap on a vertex of the $\{4,4\}$ tessellation. To ensure this remains the case for the hyperbolic tessellations, it is necessary to require that $s=4$, which is by definition a property of our subsystem hyperbolic codes. Setting $s=4$ alone is not sufficient, since we must now also ensure that the entire neighbourhood (all \textit{seven} neighbours) of each triangle operator with a given label in the lattice remains identical to that of a triangle operator with the same label in the toric code. The relative labels of these seven neighbours is determined by the homomorphism $h:G_{r,s}^{H+} \rightarrow  \mathbb{Z}_4$ defined in \Cref{eq:labelling_homomorphism}.

 Therefore, a hyperbolic tessellation is schedulable if its proper symmetry group admits the homomorphism $h$ as defined in Eq.~(\ref{eq:labelling_homomorphism}), which is the case if and only if $h(r_i)=0$ for each relation $r_i$ in the presentation of $G_{r,s}^{H+}$.
This condition is met for subsystem hyperbolic codes derived from the subset of closed $\{4c,4\}$ tessellations (where $c\in\mathbb{Z}^+$), for which the generators $g_i$ of the normal subgroup $H$ defining the compactification satisfies $h(g_i)=0$. As an example, consider the quotient group for a distance $L$ toric code which has presentation
\begin{equation}
G_{4,4}^{H+}=\langle \rho,\sigma | \rho^4=\sigma^4=(\rho\sigma)^2=(\rho\sigma^{-1})^L=(\sigma^{-1}\rho)^L=e\rangle
\end{equation}
from which it is clear that each relation $r_i$ satisfies $h(r_i)=0$. 

For schedulable subsystem hyperbolic codes, we can use the very efficient measurement schedule of Ref.~\cite{bravyi2013subsystem} (which is translationally invariant for the subsystem toric code) for each triangle operator, which requires only four time steps (one time step is the duration of a CNOT gate) to measure all $X$ and $Z$ check operators. Note that subsystem hyperbolic codes which do \textit{not} satisfy these constraints will still admit a measurement schedule, but such a schedule may be considerably less efficient and also more difficult to construct.

Given the map $m:\mathbb{Z}_4\rightarrow \mathbb{Z}_2$ defined by $m(x)=x \mod 2$ assigning a colour to a label, we see that $f(g)=m(h(g))\quad \forall g\in G_{r,s}^{H+}$, where $f$ is defined in Eq.~(\ref{eq:f_homomorphism}), and hence every schedulable code is colourable (but not vice versa, as exemplified by the $\{6,4\}$ tessellation for which $\rho^6$ is a relation yet $h(\rho^6)\neq 0$).

There is another way of interpreting the scheduling: Consider the graph which is generated by the rotation subgroup $\langle \rho, \sigma \rangle$. this group acts regularly between the triangles of the subsystem code, so there is a one-to-one map between them. The labeling is a coloring of the Cayley graph of this group (each vertex of this Cayley graph corresponds to a triangle). This coloring is achieved by a ``covering'' of the cycle graph with 4 vertices (Cayley graph of $\mathbb{Z}_4$) since this is clearly 4-colourable. More generally, we can consider normal subgroups $N$ of the group as long as this normal subgroup does not contain~$\rho$ or~$\sigma$. The number of colours in this case is the index of~$N$ in~$G$.

The dual semi-hyperbolic tessellations used for constructing the subsystem semi-hyperbolic codes in \Cref{sec:semi_hyperbolic_codes} do not have a group structure, so they cannot be labelled using the homomorphism of \Cref{eq:labelling_homomorphism} alone. However, we now show that, given a schedulable~$\{4c,4\}$ tessellation, the corresponding dual semi-hyperbolic tessellation derived from it is also schedulable. Take a schedulable $\{4c,4\}$ tessellation $V$, where we have already labelled each corner in the tessellation with an element of $\mathbb{Z}/4\mathbb{Z}$. Now consider its dual tessellation $V^*$, constructed by exchanging vertices and faces in the Hasse diagram of the tessellation~\cite{breuckmann2017homological}. Each corner in $V$ is identified by a face and vertex, and so each corner in $V$ is in one to one correspondence with a corner in $V^*$ (where the face and vertex are exchanged). We give each corner in $V^*$ the same label as the corner in $V$ that it is in one to one correspondence with. This constitutes a valid labelling of $V^*$, since each pair of corners related by $\rho$ ($\sigma$) in $V$ are related by $\sigma$ ($\rho$) in $V^*$, and $h(\rho)=h(\sigma)$ in \Cref{eq:labelling_homomorphism}. We now construct a semi-hyperbolic tessellation $V_l^*$ by tiling each face of $V^*$ with an $l\times l$ square lattice. Note that the corners of each face in $V^*$ are already labelled, so we can label $V_l^*$ just by labelling the new corners introduced by the $l\times l$ square tiling of each face. Corners related by a $\sigma$ rotation in $V^*$ are still related by a $\sigma$ rotation in $V_l^*$. Corners related by a $\rho$ rotation in $V^*$ are now related by a $(\rho\sigma^{-1})^{l-1}\rho$ translation in $V_l^*$. However, now treating $h$ as a function not a homomorphism, note that $h(\rho)=h((\rho\sigma^{-1})^{l-1}\rho)$, so the original labels retained from $V^*$ remain valid. We can therefore label the new corners in the square $l\times l$ tilings in $V_l^*$ in a way that is consistent with the corners already labelled. We now take the dual of $V_l^*$ to obtain $V_l$, preserving the labels of each corner when taking the dual as before. The tessellation $V_l$ is now used to derive a subsystem semi-hyperbolic code, and we have demonstrated that $V_l$ is schedulable if $V$ is schedulable.

\section{Subsystem semi-hyperbolic and subsystem toric code comparison}

A quantum code derived from a $\{r,s\}$-tessellation satisfies~\cite{breuckmann2016constructions}
\begin{equation}
\frac{k}{n}=1-\frac{2}{s} -\frac{2}{r} + \frac{2}{n}
\end{equation}
where $n$ is the number of physical qubits and $k$ is the number of logical qubits. A semi-hyperbolic code derived from such a code has $l^2n$ qubits, where $l$ is the dimension of the lattice tiling each face in the semi-hyperbolic code. Therefore, the number of data qubits (excluding ancillas) in a subsystem $\{8,4\}$-semi-hyperbolic code with $k$ logical qubits is $6(k-2)l^2$. To compare the performance of subsystem semi-hyperbolic codes with subsystem toric codes, we will compare each semi-hyperbolic code to multiple independent copies of a toric code with the same rate $k/n$, such that we can compare the performance keeping $k$ and $n$ fixed. Since the rate of a subsystem toric code with distance $L$ is $2/(3L^2)$, we compare our subsystem semi-hyperbolic $\{8,4\}$ codes with copies of a toric code with distance close to
\begin{equation}\label{eq:toric_vs_semi_constant_rate}
L=2l\sqrt{1-\frac{2}{k}}
\end{equation}
where $k$ is the number of qubits in the $\{8,4\}$ semi-hyperbolic code and $l$ is the dimension of the lattice tiling each face in the semi-hyperbolic code. Note that the total number of qubits including ancillas $(1+4n_a/3)n$ is proportional to the number of data qubits $n$ with the same constant of proportionality for the subsystem toric, hyperbolic and semi-hyperbolic codes.
Here, $n_a$ is the number of ancilla qubits used per triangle operator (we can always set $n_a=1$, but for some schedules setting $n_a=2$ can improve performance by parallelising the measurement schedule).
Therefore Eq.~(\ref{eq:toric_vs_semi_constant_rate}) still holds once ancillas are taken into account.

\section{Distance of subsystem hyperbolic codes}\label{sec:subsystem_hyperbolic_distance}

\begin{figure}
\includegraphics{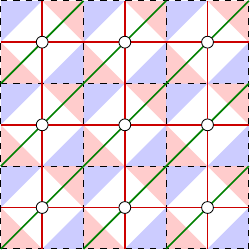}
\caption{The subsystem toric code. The black dashed lines are edges of the $\{4,4\}$ tessellation from which the subsystem toric code is derived. The edges in the $X$-type matching graph are the union of the solid red and green lines, and vertices in the matching graph are denoted by circles. Each edge in the $X$-type matching graph corresponds to a data qubit, and each face corresponds to a $Z$-type triangle operator. The solid red lines are the edges of the matching graph for the standard surface code derived from the same $\{4,4\}$ tessellation. Opposite sides are identified.}
\label{fig:subsystem_toric_matching_graph}
\end{figure}

We can determine the distance of the subsystem hyperbolic and semi-hyperbolic codes by considering their matching graphs. Each vertex in the $X$-type matching graph corresponds to an $X$ stabiliser, and there is an edge between each pair of stabilisers $u$ and $v$ for which a single $Z$ error on a data qubit anti-commutes with both $u$ and $v$. Each face in the $X$-type matching graph corresponds to a $Z$-type triangle operator. Each non-contractible closed loop in the $X$-type matching graph corresponds to a logical $Z$ operator. Therefore, the $Z$-distance of the code is determined by the shortest non-contractible closed loop in the $X$-type matching graph. A $Z$-type matching graph can be defined analogously for $Z$-type stabilisers and so the $X$ distance of the code is determined by the shortest non-contractible closed loop in the $Z$-type matching graph.

For the subsystem toric, hyperbolic and semi-hyperbolic codes we construct, the $X$-type matching graph is isomorphic to the $Z$-type matching graph, since the $Z$-type matching graph can be obtained from the $X$-type matching graph (and vice versa) by a single rotation that is also a symmetry of the tessellation from which the code is derived. Therefore, the $Z$ and $X$ distances are the same for these codes.

\begin{figure}
\includegraphics[width=0.8\columnwidth]{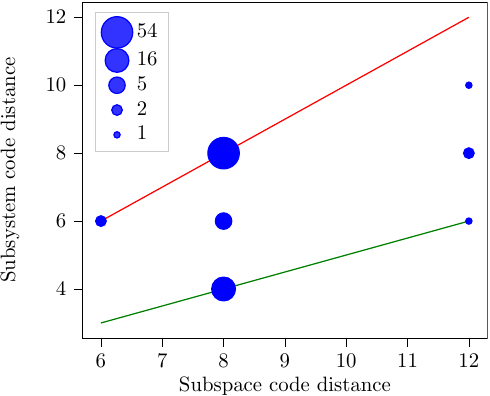}
\caption{For all $l=2$ $\{8,4\}$ subsystem semi-hyperbolic codes we constructed, here we plot the distance of each code ($y$-axis) against the distance of the (subspace) semi-hyperbolic surface code derived from the same tessellation ($x$-axis), computed using the method in Ref.~\cite{erickson2005greedy}. The size of each blue circle corresponds to the number of codes we found with the same ($x$,$y$) coordinate on the figure, and the number of codes for each size of circle is given in the legend.}
\label{fig:subsystem_vs_subspace_distances}
\end{figure}

We will now consider how the distance of a subsystem hyperbolic or semi-hyperbolic code compares to the distance of the subspace CSS (surface) code derived from the same tessellation. To do so, we will consider the structure of the matching graph for both codes. The solid red lines in \Cref{fig:subsystem_toric_matching_graph} form the edges of the $Z$-type matching graph for the toric code, and so the length of the shortest non-contractible loop in that graph is the $X$ distance of the toric code. We can obtain the $X$-type matching graph for the \textit{subsystem} toric code derived from the same tessellation by adding in the green edges, also shown in \Cref{fig:subsystem_toric_matching_graph}, and keeping the same set of vertices. Each green edge in the subsystem toric code $X$-type matching graph is equivalent (up to a triangle operator) to a \textit{pair} of red edges. 
Therefore, the distance between two vertices in the matching graph consisting only of red edges can \textit{at most} be reduced by half by the inclusion of the green edges (and inclusion of the green edges cannot increase the distance between vertices). 

For the subsystem hyperbolic and semi-hyperbolic codes, we again find that both the $Z$-type and $X$-type matching graphs can be constructed by adding additional edges to the $Z$-type matching graph $V_Z$ of the subspace codes derived from the tessellation, where each of these additional edges is equivalent to a pair of edges in~$V_Z$. 
Therefore, the shortest non-contractible loop in either the $Z$-type or $X$-type matching graph for a subsystem hyperbolic or semi-hyperbolic code is between one and two times smaller than the shortest non-contractible loop in the $Z$-type matching graph of the subspace code derived from the same tessellation. Consequently, given a hyperbolic or semi-hyperbolic code with $X$ distance~$d_X$, the distance $d$ of the subsystem hyperbolic or semi-hyperbolic code derived from the same tessellation is bounded by $d_X/2\leq d\leq d_X$. Furthermore, the $X$ distance of hyperbolic codes we consider is always less than or equal to their $Z$ distance. Both the subsystem toric code and standard toric code have distance $d=L$, but for the subsystem hyperbolic and semi-hyperbolic codes we construct, the subsystem codes do have a reduced distance compared to surface codes derived from the same tessellation. This is shown in \Cref{fig:subsystem_vs_subspace_distances}, which compares the distance of $l=2$, $\{8,4\}$ subsystem semi-hyperbolic codes to the distance of the subspace semi-hyperbolic codes derived from the same tessellations. We see that the distance of each subsystem code can be reduced by up to $2\times$ relative to the subspace code derived from the same tessellation as expected, with some subsystem codes not suffering any reduction in distance.

\section{The Construction for General~LDPC~Quantum~Codes}\label{app:general_construction}
	The ideas from \Cref*{sec:2d_construction,sec:subsystem_hyperbolic_codes} of the main text readily apply to the more general class of CSS stabilizer codes.
	In a CSS stabilizer code the stabilizer checks operate exclusively as either Pauli-$X$ or Pauli-$Z$ on all of the qubits in its support.
	The \emph{Tanner graph} associated to a CSS code is the graph with vertices corresponding to qubits, $X$-checks and $Z$-checks.
	There is an edge between two vertices if and only if one vertex corresponds to a check and the other to a qubit in its support.
	In order to define the merging and splitting for a CSS LDPC code, let us pick a $Z$-check $s_Z$ and consider the subgraph~$T'$ of the Tanner graph consisting of: all qubits in the support of $s_Z$, their connected $X$-checks, as well as all edges connecting them.
	We call a set of $X$-checks in $T'$ a \emph{local cut-set} if removing them and their incident edges from $T'$ renders it disconnected.
	We call a local cut-set \emph{independent} if the checks contained are linearly independent.
	Let $\mathcal{I}$ be the labels of a local cut-set.
	The checks belonging to $\mathcal{I}$ are \emph{merged} by defining a new Tanner graph in which all of the vertices of $\mathcal{I}$ are identified.
	This procedure is also known as \emph{vertex contraction} in the graph theory literature.
	
	Merging an independent, local cut-set reduces the number of $X$-checks by $|\mathcal{I}|-1$.
	Since the number of physical qubits was not changed and the checks were independent there must be $|\mathcal{I}|-1$ new logical degrees of freedom.
	However, the operator algebra of these degrees of freedom will be supported on at most $|\text{supp}(s_Z)| \in O(1)$ physical qubits.
	Therefore they do not offer protection and we will consider them to act on gauge qubits.
	
	\begin{figure}
		\includegraphics[width=0.95\columnwidth]{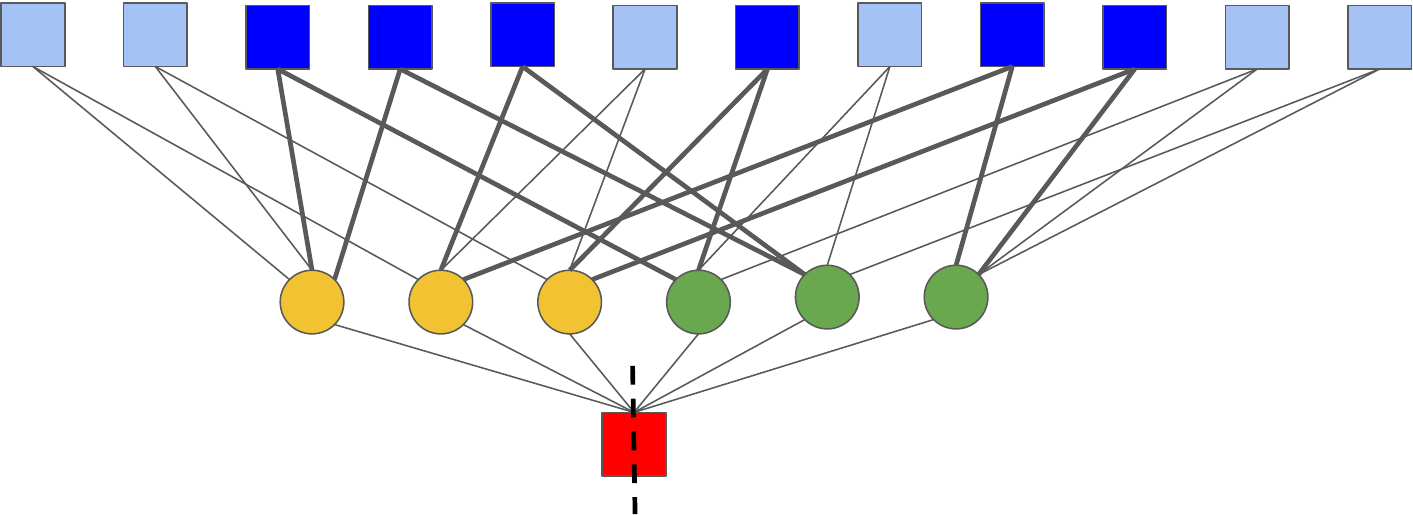}
		\caption{Neighborhood of a $Z$ stabilizer check in the Tanner graph of a CSS quantum code. Circles represent qubits, blue boxes in the top row represent $X$ checks which are in the neighborhood of a $Z$ check (red box at the bottom). Bold lines represent a cut set which induces a partition of the qubits into two sets (yellow and green) and  cut on the $Z$ check. $X$ checks in dark blue belong to the set $\mathcal{I}$ and will be merged.}
		\label{fig:LDPC_subsystem}
	\end{figure}
	
	We will now describe a basis for the operator algebra acting on the gauge qubits.
	Let $C_1,\dotsc,C_l$ be the qubits belonging to the connected components of $T'$ induced by the cut-set.
	We define $Z$-type operators~$Z^g_{i}$ which act on all qubits in~$C_i$.
	Note that each $Z^g_{i}$ commutes with all $X$-checks in the code, because  all $X$-checks not belonging to the local cut set must overlap with~$s_Z$ on either the empty set or any of the~$C_i$.
	Since all $X$-checks commute with~$s_Z$ they must have even support on~$C_i$ and hence commute with $Z^g_{i}$.
	A set of operators which anti-commute with the~$Z^g_{i}$ are the $X$-checks in the local cut-set.
	
	The merging and splitting procedure reduces the number of linearly independent stabilizers~$r$ and increases the number of gauge operators~$g$.
	The number of physical qubits~$n$ and logical qubits~$k$ are unaffected, so that \Cref{eqn:parameter_dependency} is satisfied.
	
	We note that this procedure will generally affect the distance, as it does for the suface, toric and hyperbolic codes.
	An extreme example is the surface code defined on a square lattice, where merging the top left and bottom right $X$ checks of each face ($Z$ check) leads to the code distance turning constant.
	Demonstrating that the procedure gives a subsystem code with growing distance therefore has to be informed by the structure of the code.

\section{Scheduling from group homomorphisms}
	
	In \Cref{app:labelling_homomorphism} we showed that an efficient syndrome measurement schedule for subsystem hyperbolic codes could be constructed if the orientation-preserving symmetry group $G_{r,s}^{H+}$ of the tessellation (generated by rotations $\rho$ and $\sigma$) admits a homomorphism $f:G_{r,s}^{H+}\rightarrow \mathbb{Z}_4$ to the cyclic group $\mathbb{Z}_4$, with $f$ defined by $f(\rho)=f(\sigma)=1$. This homomorphism is a useful tool for scheduling subsystem hyperbolic codes for the same reason that translation invariance is useful for scheduling Euclidean surface codes: the problem of scheduling the entire code reduces to the problem of scheduling only a small number of stabilisers in a region of the tessellation.
	
	While the homomorphism $f:G_{r,s}^{H+}\rightarrow \mathbb{Z}_4$ is a useful tool for scheduling the subsystem hyperbolic codes, such a homomorphism only exists for a subset of $\{r,s\}$ tessellations (for which four divides both $r$ and $s$). In this section we will look for homomorphisms from $G_{r,s}^{H+}$ to \textit{any} cyclic group, in the hope that these homomorphisms will be a useful tool for scheduling subspace hyperbolic codes based on a wider range of tessellations, where each $Z$ stabiliser (plaquette) and $X$ stabiliser (site) is measured using a circuit with a single ancilla qubit. Each corner $C_g$ of a face of the tessellation is identified with an element $g\in G_{r,s}^{H+}$. By finding a homomorphism $f:G_{r,s}^{H+}\rightarrow \mathbb{Z}_n$ to a cyclic group $\mathbb{Z}_n$, we can label each corner uniquely with an element in $\mathbb{Z}_n$. The function $f$ is a homomorphism if and only if $f(r_i)=0$ for each relation $r_i$ in the presentation of $G_{r,s}^{H+}$. The tessellation group $G_{r,s}^{H+}$ has presentation
	\begin{equation}
	    G_{r,s}^{H+}\coloneqq \langle \rho,\sigma | (\rho\sigma)^2=\rho^r=\sigma^s=e \rangle
	\end{equation}
	from which we see that $(\rho\sigma)^2$ is always a relation, and hence $f$ must always satisfy $f((\rho\sigma)^2)=0$.
	
	For the homomorphism $f:G_{r,s}^{H+}\rightarrow \mathbb{Z}_n$ to be useful for scheduling, we will require that it must satisfy a additional properties. Firstly, the homomorphism should not be defined by $f(\rho)=f(\sigma)=0$, since this homomorphism does not give us any additional information. Secondly, the label of each corner~$C_g$ should be different to the corner $C_{g\rho\sigma}$. This is because~$C_g$ and~$C_{g\rho\sigma}$ overlap on an edge~$e$ in such a way that, if both corners had the same schedule, two CNOT gates applied to the qubit at~$e$ would occupy the same time step. 
	
	We will assume that can have more than one ancilla for each stabiliser, to parallelise the measurement circuits. If we were instead to insist that only a single ancilla be used, then we must require that all corners belonging to the same vertex must have different labels. This is because these corners share an ancilla qubit on the vertex, and two CNOT gates cannot be applied to the ancilla qubit within the same time step. Furthermore, we would also require that all corners belonging to a face must have a different label, since only a single CNOT gate can be applied to the ancilla qubit in the centre of each face in each time step.
	
	Therefore for each tessellation $\{r,s\}$, we will seek to find a cyclic group order $n$ and elements $x,y\in\mathbb{Z}_n$ such that the function defined by $f(\rho)=x, f(\sigma)=y$ extends to a homomorphism $f:G_{r,s}^{H+}\rightarrow \mathbb{Z}_n$. The restrictions on the relations in the presentation of $G_{r,s}^{H+}$, along with the additional three properties we have imposed, correspond to the following constraints on $x,y,n$:
	\begin{align}\label{eq:cyclic_homomorphism_constraints}
	\begin{split}
	    rx &= 0 \mod n \\
	    sy &= 0 \mod n \\
	    2(x+y) &= 0 \mod n \\
	    x + y &\neq 0 \mod n
	\end{split}
	\end{align}
and if we could use only a single ancilla per stabiliser, then we would additionally have the constraints
\begin{align}
\begin{split}
\mathrm{lcm}(x,n) &= rx \\
\mathrm{lcm}(y,n) &= sy.
\end{split}
\end{align}
	
For all $r,s\leq10$ we have searched for all $n,x,y$ satisfying Eq.~\ref{eq:cyclic_homomorphism_constraints} (for $n<5\max(r,s)$) and list all the tessellations we found which admitted at least one such homomorphism in Table~\ref{table:homomorphisms}.

\begin{table}
\begin{tabular}{ccccc}
$r$ & $s$ & $n$ & $x$ & $y$ \\
\hline
3&6&6&2&1 \\
4&4&4&1&1 \\
4&8&4&1&1 \\
5&10&10&2&3 \\
6&6&6&1&2 \\
6&9&6&1&2 \\
8&8&8&1&3 \\
10&10&10&1&4 
\end{tabular}
\caption{Solutions to Eq.~\ref{eq:cyclic_homomorphism_constraints} for all $r,s\leq10, r\leq s$. By symmetry, solutions for $r\geq s$ can be found by exchanging column $r$ with $s$ and column $x$ with $y$. For each tessellation $\{r,s\}$, we give the parameters $n,x,y$ defining only one homomorphisms $f:G_{r,s}^{H+}\rightarrow Z_n$ (the homomorphism which minimimises both $n$ and $x$). There are at least two solutions for each tessellation.}\label{table:homomorphisms}
\end{table}

While we have found homomorphisms to cyclic groups for many tessellations, we did not find any for the $\{5,5\}$ code, which has the desirable properties of being self-dual and having low stabiliser weights. Therefore, an interesting question is whether there exist homomorphisms to groups that are not cyclic, and which contain a small number of elements, but otherwise satisfy the constraints of \Cref{eq:cyclic_homomorphism_constraints}. If such a homomorphism exists for tessellations such as $\{4,5\}$ and $\{5,5\}$, the trade off of circuit-level threshold and encoding rate for these codes may be very favourable.

\section{Additional numerical results}\label{app:thresholds}

\begin{table}
\begin{tabular}{ccc}
Schedule & $p^{th}_{depol}$ & $p_{depol}^{th,*}$ \\
\hline
ZX & 0.666(1) &  0.666(1)  \\
$Z^2X^2$ & 0.757(1) & 0.6587(9)  \\
$Z^3X^3$ & 0.810(2) & 0.676(1)  \\
$Z^4X^4$ & 0.811(2) & 0.669(2) \\
$Z^5X^5$ & 0.792(2) & 0.652(2) \\
$Z^{10}X^{10}$ & 0.522(2) & 0.493(1) \\
\end{tabular}
\caption{Thresholds (in $\%$) for the subsystem toric code for some balanced homogeneous schedules under the circuit-level depolarising noise model, each computed using the critical exponent method of Ref.~\cite{wang2003confinement} to analyse results from Monte Carlo simulations using subsystem toric codes with distances $L=26,30,34,38,42,46$. Numbers in brackets are the $1\sigma$ uncertainties in the last digit. For each threshold, we keep the number of syndrome extraction rounds constant for all codes, always using at least 92 rounds to ensure boundary effects (in time) are small even for the largest codes. For the column with an asterisk, gauge fixing was not used when decoding.}\label{table:toric_depolarising_thresholds}
\end{table}

\begin{table}
\begin{tabular}{ccccc}
Schedule & $p_X^{th}$ & $p_X^{th,*}$ & $p_Z^{th}$ & $p_Z^{th,*}$ \\
\hline
ZX & 0.515(1) & & 0.515(1) &  \\
$Z^2X^2$ & 0.5863(9) & & 0.5863(9) & \\
$Z^3X^3$ & 0.628(1) && 0.628(1) & \\
$Z^4X^4$ & 0.631(2) && 0.631(2)  & \\
$Z^5X^5$ & 0.619(2) && 0.619(2)  & \\
$ZX^2$ &0.3928(8) & 0.3928(8) & 0.749(1) & 0.625(3) \\
$ZX^3$ & 0.3236(9) & 0.3236(9) & 0.931(1) & 0.7234(9) \\
$ZX^5$ & 0.2449(5) & 0.2449(5) & 1.160(2) & 0.816(2) \\
$ZX^{10}$ & 0.1595(4) & 0.1595(4) & 1.430(3) & 0.902(2) \\
$Z^2X^{10}$ & 0.2394(5) & 0.2259(5) & 1.197(3) & 0.821(2)  \\
$X$ & 0 &0& 2.2231(1) & 1.029(2) \\
\end{tabular}
\caption{Thresholds (in $\%$) for the subsystem toric code for various homogeneous schedules under the independent circuit-level noise model, each computed using the critical exponent method of Ref.~\cite{wang2003confinement} to analyse results from Monte Carlo simulations using subsystem toric codes with distances $L=26,30,34,38,42,46$. Numbers in brackets are the $1\sigma$ uncertainties in the last digit. For each threshold, we keep the number of syndrome extraction rounds constant for all codes, always using at least 92 rounds to ensure boundary effects (in time) are small even for the largest codes. For the final two columns (with asterisks in the title), gauge fixing was not used even when possible.}\label{table:toric_independent_thresholds}
\end{table}

\begin{figure}
\includegraphics[width=0.7\columnwidth]{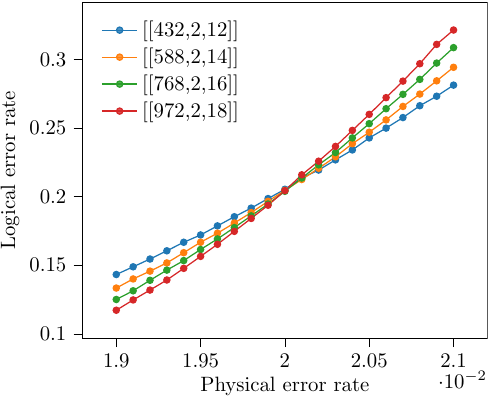}
\caption{Subsystem toric code threshold with a phenomenological noise model, and without using gauge fixing (triangular lattice matching graph). Using the critical exponent method we find a threshold of $0.02004(2)$.}
\label{fig:phenomenological_triangular}
\end{figure}

\begin{figure}
\includegraphics[width=0.7\columnwidth]{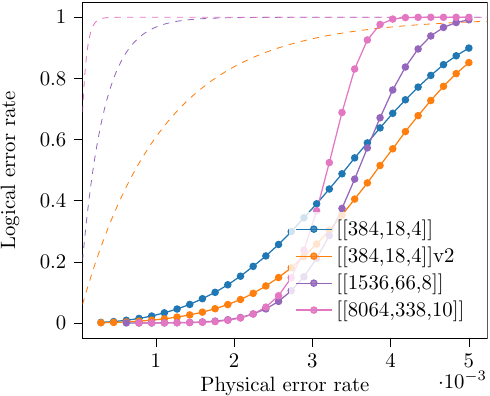}
    \caption{Performance of the extremal subsystem $\{8,4\}$ $l=2$ semi-hyperbolic codes under a circuit-level depolarising noise model. A homogeneous $(ZX)^{20}$ schedule is used for all codes, and the $y$ axis is the probability that at least one logical $Z$ error occurs. Dashed lines are the probability of a $Z$ error occurring on at least one of $k$ physical qubits without error correction under the same error model and for the same duration (80 time steps), with $k=4$ (orange), $k=8$ (purple) and $k=10$ (pink).}
    \label{fig:semi_hyperbolic_l2}
\end{figure}

\begin{figure}
\includegraphics[width=0.8\columnwidth]{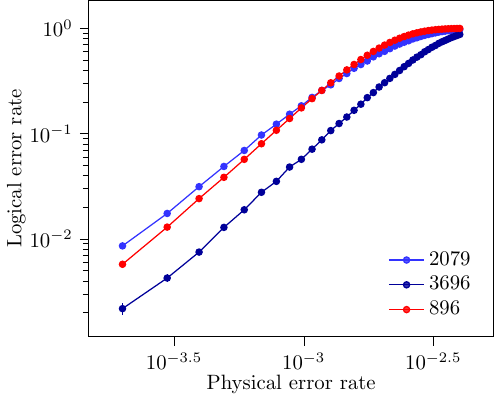}
\caption{Performance of a~[[384,66,4]]~$\{8,4\}$ subsystem hyperbolic code (red) compared to the $L=3$ and $L=4$ subsystem toric codes (shades of blue) using a $(ZX)^{10}$ schedule with the circuit-level depolarising error model. We use 33 independent copies of the subsystem toric codes to fix the number of logical qubits at $k=66$. In the legend we give the number of physical qubits used, including ancillas.}
\label{fig:small_subsystem_hyperbolic_vs_subsystem_toric}
\end{figure}

In this section we give some additional numerical results from simulations of the subsystem toric and semi-hyperbolic codes. In \Cref{table:toric_depolarising_thresholds}, we give thresholds for the subsystem toric code code under a circuit-level depolarising noise model using gauge fixing with balanced schedules. In \Cref{table:toric_independent_thresholds}, we give thresholds for the subsystem toric code under an independent circuit-level noise model using both balanced and unbalanced schedules. 
In \Cref{fig:phenomenological_triangular} we plot the threshold for the subsystem surface code with a phenomenological noise model, which we find to be $0.02004(2)$ using the critical exponent method of Ref.~\cite{wang2003confinement}. In \Cref{fig:semi_hyperbolic_l2} we plot the threshold of the~$l=2$~$\{8,4\}$ subsystem semi-hyperbolic codes \textit{without} adjusting for the number of logical qubits, unlike in the text. This is helpful to better understand the logical error rates of the codes themselves, but less so for understanding the threshold for the logical error rate per logical qubit, for which multiple independent copies of the smaller codes should be taken, as done in the main text. In \Cref{fig:small_subsystem_hyperbolic_vs_subsystem_toric} we compare the performance of a~[[384,66,4]]~$\{8,4\}$ subsystem hyperbolic code with 33 copies of $L=3$ and $L=4$ subsystem toric codes, all encoding 66 logical qubits. Since this hyperbolic code is quite small, its overhead $n/(kd^2)\approx 0.36$ is less favourable than that of the much larger [[8064,338,10]] subsystem semi-hyperbolic code analysed in \Cref*{sec:subsystem_semi_hyperbolic_performance} of the main text, for which $n/(kd^2)\approx 0.24$. However, as can be seen from \Cref{fig:small_subsystem_hyperbolic_vs_subsystem_toric}, the [[384,66,4]] subsystem hyperbolic code still uses $2.3\times$ fewer physical qubits than the subsystem toric code to achieve the same logical error rate per logical qubit below a circuit-level depolarising physical error rate of $0.1\%$. Furthermore, it only requires 896 physical qubits to implement this subsystem hyperbolic code including ancillas, compared to the $18,816$ needed for the [[8064,338,10]] subsystem semi-hyperbolic code.

\end{document}